\pdfoutput=1
\documentclass{aa}

\usepackage{txfonts}
\usepackage{natbib}
\usepackage[colorlinks=true,linkcolor=blue,citecolor=blue]{hyperref}

\begin{document}

\title{Accelerated particle beams in a 3D simulation of the quiet Sun}


\author{
    L. Frogner
    \and
    B. V. Gudiksen
    \and
    H. Bakke
}

\institute{
   Institute of Theoretical Astrophysics,
   University of Oslo,
   P.O.Box 1029 Blindern,
   N-0315 Oslo,
   Norway
   \and
   Rosseland Centre for Solar physics (RoCS),
   University of Oslo,
   P.O.Box 1029 Blindern,
   N-0315 Oslo,
   Norway
}

\date{}

\abstract
{
    Observational and theoretical evidence suggest that beams of accelerated particles are produced in flaring events of all sizes in the solar atmosphere, from X-class flares to nanoflares. Current models of these types of particles in flaring loops assume an isolated 1D atmosphere.
}
{
    A more realistic environment for modelling accelerated particles can be provided by 3D radiative magnetohydrodynamics codes. Here, we present a simple model for particle acceleration and propagation in the context of a 3D simulation of the quiet solar atmosphere, spanning from the convection zone to the corona. We then examine the additional transport of energy introduced by the particle beams.
}
{
    The locations of particle acceleration associated with magnetic reconnection were identified by detecting changes in magnetic topology. At each location, the parameters of the accelerated particle distribution were estimated from local conditions. The particle distributions were then propagated along the magnetic field, and the energy deposition due to Coulomb collisions with the ambient plasma was computed.
}
{
    We find that particle beams originate in extended acceleration regions that are distributed across the corona. Upon reaching the transition region, they converge and produce strands of intense heating that penetrate the chromosphere. Within these strands, beam heating consistently dominates conductive heating below the bottom of the transition region. This indicates that particle beams qualitatively alter the energy transport even outside of active regions.
}
{}


\maketitle


\section{Introduction}
The solar atmosphere is full of energetic particles. They are produced when ambient plasma particles are accelerated out of thermal equilibrium by a strong electric field or by rebounding off moving magnetic elements. Upon leaving the site of acceleration, their available trajectories are limited by the Lorentz force, which compels charged particles to follow the direction of the magnetic field. Consequently, the accelerated particles form coherent beams. The interactions of these beams with the ambient plasma are believed to be a key mechanism in solar flares.

It is generally accepted that flares are powered by the relaxation of stresses in the magnetic field through the process of magnetic reconnection. The release of magnetic energy manifests as electric field enhancements and plasma flows, leading to strong currents with associated resistive heating as well as jets of outflowing plasma and magnetoacoustic waves. This also creates conditions favourable for particle acceleration. The ensuing beams of energetic particles, which may account for a significant portion of the flare energy \citep{Lin1971}, transfer energy to the plasma along their trajectories through Coulomb interactions. The X-ray bremsstrahlung emitted in these interactions can escape the atmosphere relatively unaffected, and thus it provides valuable information about the energy distribution of the particles and the plasma conditions at the site of emission.

Signatures of accelerated particles, in particular non-thermal electrons, are found in observed hard X-ray spectra from active region flaring events, ranging from large flares with energies up to $10^{32}$ erg \citep[see][for a review]{Benz2017} to $10^{27}$ erg microflares at the sensitivity limit of current instruments \citep[e.g.][]{Christe2008}. This suggests that particle beams play an active role in flares of all sizes. Beyond hard X-ray detectability, \citet{Parker1988} predicted frequent impulsive heating events with energies of the order of $10^{24}$ erg that are associated with small-scale reconnection due to the continuous interweaving of the magnetic field by photospheric convective motions. Signs of these types of events, dubbed nanoflares, have been observed down to $10^{25}$ erg as ultraviolet (UV) and soft X-ray flashes in the chromosphere and transition region of the quiet Sun \citep{Krucker1998, Benz2002}. Based on detailed 1D simulations, \citet{Testa2014} found that non-thermal electron beams were required to reproduce the UV spectra in their observations of nanoflares.

Early models of particle beams were mostly based on simple analytical expressions for the mean collisional change in velocity of energetic particles moving through the atmospheric plasma \citep{Brown1972, Syrovatskii1972, Emslie1978}. The response of the atmosphere to an injected beam of non-thermal electrons has been studied by incorporating these expressions into the energy equation of 1D hydrodynamics simulations \citep{Somov1981, MacNeice1984, Nagai1984}. More recently, the realism of these types of simulations was improved significantly by the inclusion of detailed radiative transfer \citep{Hawley1994, Abbett1999, Allred2005}. A more general treatment of the accelerated particles is possible by numerically solving the Fokker--Planck equation governing the evolution of the particle distribution. Due to high computational demand, this method was, initially, only used to study the detailed propagation and bremsstrahlung emission of non-thermal electrons in simple static model atmospheres \citep{Leach1981, Leach1983}. However, in state-of-the-art 1D flare simulations \citep{Liu2009, Allred2015, RubiodaCosta2015}, it has now largely replaced the approximate heating expressions derived from mean scattering theory. The high level of detail in these simulations makes them a powerful tool for studying flare dynamics and for generating synthetic diagnostics of flaring atmospheres.

Yet, by nature of their dimensionality, simulations of this kind can only consider a single flaring loop at a time, and these loops do not live in isolation. They are part of a continuous magnetic field embedded in a 3D plasma environment. Magnetic reconnection and associated acceleration of energetic particle beams are driven by the overall evolution of the atmosphere, which in turn is influenced by the collective interaction of the beams with the ambient plasma. With the drastic increase in computing power along with the advent of advanced 3D radiative magnetohydrodynamics (MHD) codes in the past couple of decades \citep{Voegler2005, Felipe2010, Gudiksen2011}, realistic simulations that self-consistently reproduce the overall structure and evolution of diverse features of the solar atmosphere are now possible. Incorporating acceleration and propagation of energetic particle beams into these types of 3D simulations would greatly benefit our understanding of the role of particle beams on the Sun.

We have taken the first step towards this goal, and here present a simple treatment of energy transport by accelerated particles applied to a realistic 3D simulation of the quiet solar atmosphere. This work is a further development of the model introduced in \citet{Bakke2018}. A related approach was recently used by \citet{Ruan2020} to incorporate electron beams into a 2.5D MHD simulation of a large flare.

A brief description of the radiative MHD code that we employ is given in Sect. \ref{sec:atmospheric_simulation}, where we also present the simulated atmosphere used for this paper. In Sect. \ref{sec:accelerated_particles}, we present the inclusion of accelerated particle beams, starting with the method of detecting reconnection sites, followed by the acceleration model and the particle transport model. Methods for reducing the computational demand, as well as for selecting values for the free parameters, are respectively discussed in Sects. \ref{sec:tuning} and \ref{sec:effect_p_delta}. Section \ref{sec:results} contains our results for the transport of energy by particle beams. These are discussed in Sect. \ref{sec:discussion}, where we also consider future work.

\section{Methods}

\subsection{Atmospheric simulation}
\label{sec:atmospheric_simulation}
We used the Bifrost code \citep{Gudiksen2011} for simulating a 3D region of the upper solar atmosphere, spanning from the top of the convection zone to the corona. Bifrost solves the resistive MHD equations with the inclusion of radiative transfer and field-aligned thermal conduction. The equation of state is computed under the assumption of local thermodynamic equilibrium (LTE), using the Uppsala Opacity Package \citep{Gustafsson1975}. The radiative transfer computation encompasses optically thin emission from the upper chromosphere and corona, approximated non-LTE radiative losses from chromospheric hydrogen, calcium and magnesium \citep{Carlsson2012}, and full radiative transfer with scattering and LTE plasma opacities in the photosphere and convection zone \citep{Hayek2010}. To maintain numerical stability, the code employs a slightly enhanced overall diffusivity in combination with spatially adaptive local diffusion (so-called hyper diffusion).

For this paper, the atmospheric environment for the accelerated particles was provided by a horizontally periodic Bifrost simulation of a $24 \times 24\;\mathrm{Mm}$ patch of the atmosphere that spans vertically from $2.5\;\mathrm{Mm}$ below the photosphere to $14.3\;\mathrm{Mm}$ above it. The simulation has a resolution of 768 grid cells along each dimension, with a uniform grid cell extent of $31\;\mathrm{km}$ in the horizontal directions. Along the vertical direction, the grid cell extent is about $12\;\mathrm{km}$ in the layer between the photosphere (at height zero) and the height of $4\;\mathrm{Mm}$, in order to resolve the abrupt local variations near the transition region. Away from this layer, the extent increases evenly in both directions to about $21\;\mathrm{km}$ near the bottom of the simulation box and $80\;\mathrm{km}$ near the top.

Convective motions are maintained in the sub-photospheric part of the simulation by injection of heat through the bottom boundary, balanced by radiative cooling in the photosphere. These motions lead to acoustic shocks and braiding of magnetic field lines, which produce a hot chromosphere and corona.

The corona was initially configured with an ambient magnetic field roughly oriented along the $x$-direction. A magnetic flux emergence scenario was then incited by injecting a 2000 G $y$-directed magnetic field, covering $x \in [4, 18]$ Mm and the full extent in $y$, through the bottom boundary. The flux sheet was broken up by convective motions as it rose to the photosphere. Here, the strongest concentrations of the field broke through and expanded into the upper atmosphere, carrying with it cool photospheric plasma. The expanding magnetic bubbles were eventually confined by the ambient coronal field. Interactions between these bubbles and with the ambient coronal field lead to magnetic reconnection at various heights and ensuing explosive events such as Ellerman bombs and UV bursts. See \citet{Hansteen2019} for a more detailed description and analysis of the simulation. It should be noted that all flaring events produced in the simulation are small (at most $\sim 10^{25}$ erg), and can generally be characterised as nanoflares.

\begin{figure}[!thb]
    \includegraphics{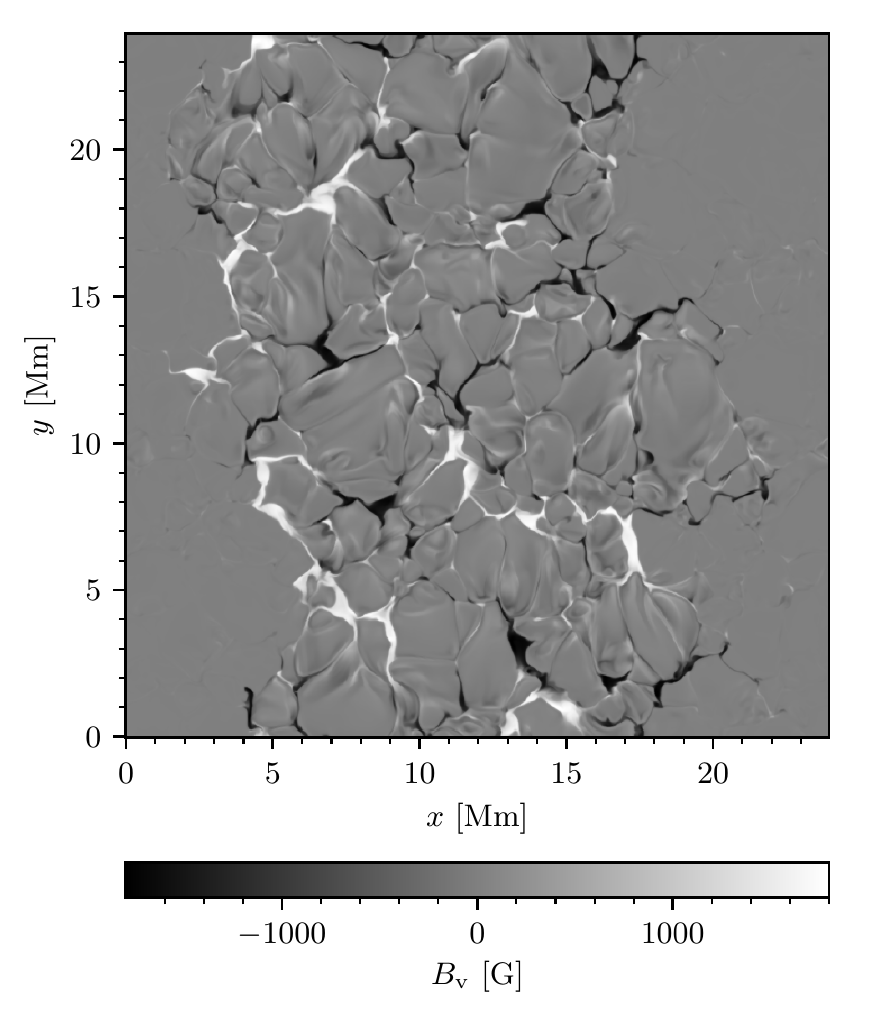}
    \centering
    \caption{Vertical magnetic field $B_\mathrm{v}$ in the photosphere (height zero) of the simulation snapshot.}
    \label{fig:magnetogram}
\end{figure}
\begin{figure}[!thb]
    \includegraphics{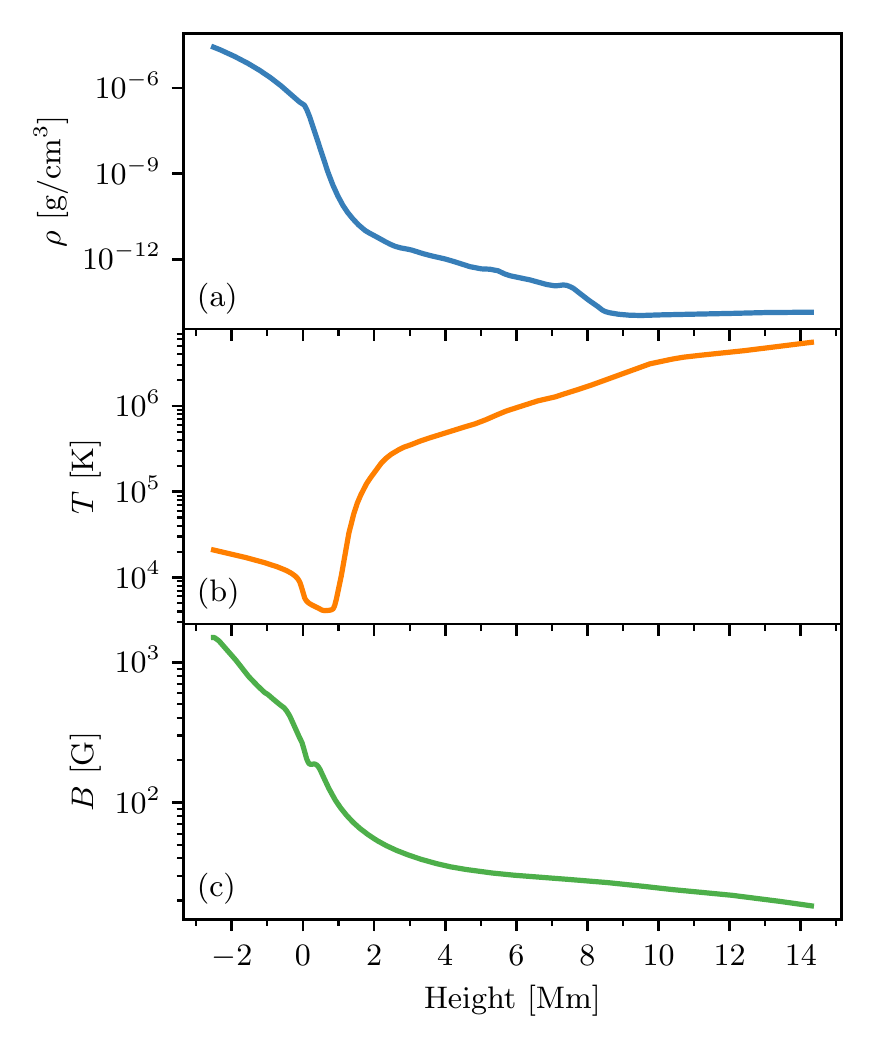}
    \centering
    \caption{Horizontally averaged mass density $\rho$ (panel (a)), temperature $T$ (panel (b)) and magnetic field strength $B$ (panel (c)) as a function of height in the simulation snapshot.}
    \label{fig:atmospheric_height_profiles}
\end{figure}
This paper considers a snapshot of the atmosphere at a single instant of the continuously evolving dynamic simulation, 8220 s after the magnetic flux sheet was injected. Figure \ref{fig:magnetogram} shows the vertical component of the photospheric magnetic field at this time. The variation in horizontally averaged mass density, temperature, and magnetic field strength with height are shown in Fig. \ref{fig:atmospheric_height_profiles}. In this figure, the presence of the relatively dense and cool magnetic bubbles filling parts of the corona is apparent in the density and temperature profiles. The noticeable break in the density profile near the height of 8 Mm corresponds to the top of the main bubble.

\subsection{Accelerated particles}
\label{sec:accelerated_particles}
Energetic electrons and ions are produced through various acceleration mechanisms during magnetic reconnection. Due to the Lorentz force, the particles follow a gyrating trajectory around the magnetic field as they travel away from the reconnection site. At the same time, they exchange energy and momentum with the background plasma through Coulomb collisions.

Based on these processes, we developed a model for the production and transport of accelerated particles suitable for integration into a 3D MHD simulation. The first step was to identify the grid cells of the simulation domain occupying locations with magnetic reconnection. In each of these grid cells, the energy distribution of the locally accelerated particles was estimated. Finally, the heating of the ambient plasma by the passing energetic particle beam was computed along the length of the magnetic field line going through the centre of each grid cell. The following sections describe the steps of this model in detail.

\subsubsection{Reconnection sites}
\label{sec:reconnection_sites}
It is well established that particle acceleration is associated with magnetic reconnection. Reconnection takes place where regions of opposite magnetic polarity come together and produce a strong rotation of the magnetic field. A magnetic diffusion region arises around the interface between the two reconnecting magnetic domains, where the gradients are strong enough to break the coupling between the magnetic field and the plasma. Inside this diffusion region, free magnetic energy is released in several different ways. The electric field induced by the rotation of the magnetic field creates a thin layer of strong current, which heats the local plasma through Joule heating. In addition, plasma is propelled away from the reconnection site from the ends of the current sheet by the magnetic tension force. Finally, a fraction of the local charged particles are accelerated to very high energies, as we discuss further in Sect. \ref{sec:initial_particle_distributions}.

\begin{figure*}[!thb]
    \includegraphics{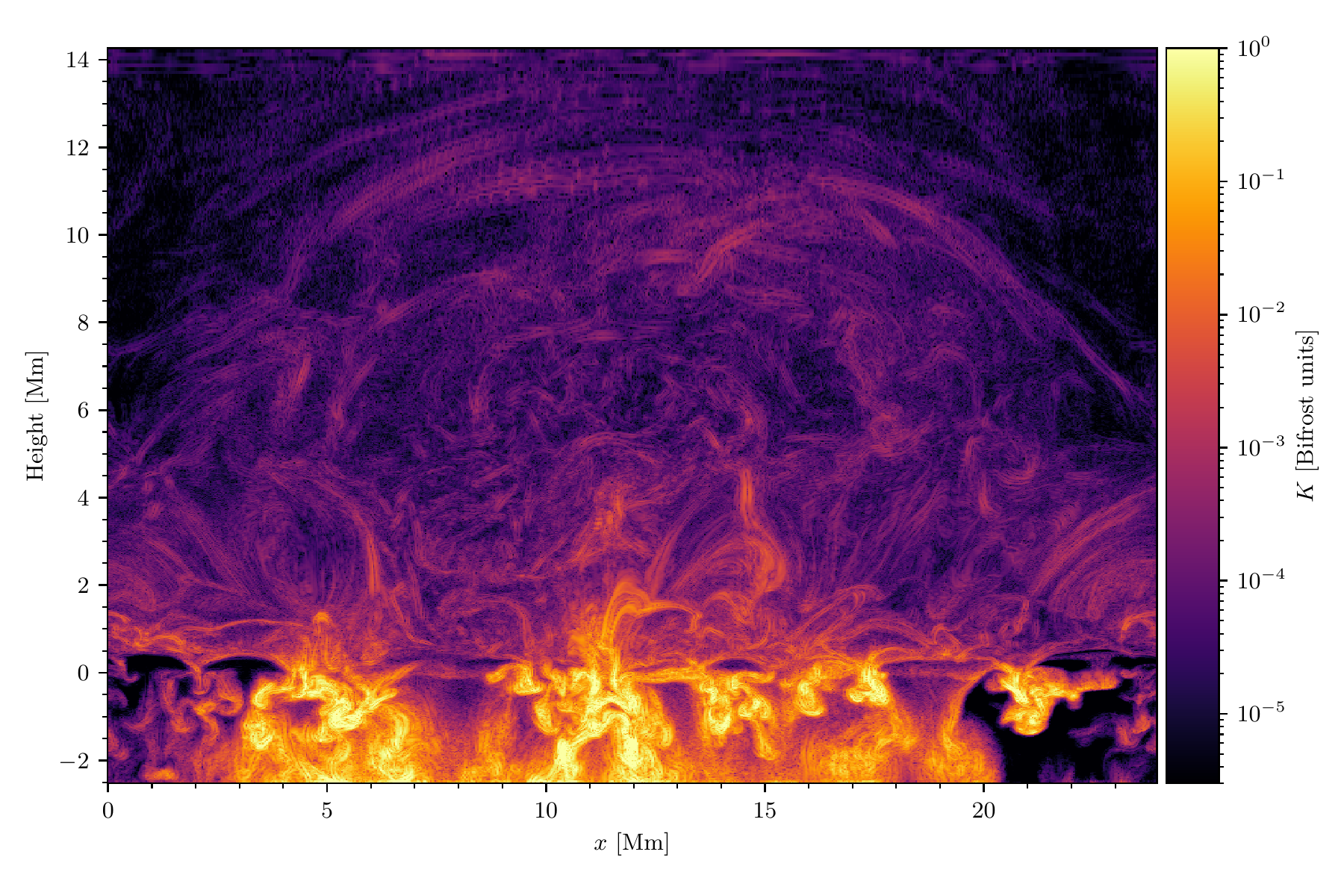}
    \centering
    \caption{Values of the reconnection factor $K$ (Eq. \eqref{eq:krec}) in a slice through the $y$-axis of the simulation snapshot, at $y = 10.67\;\mathrm{Mm}$.}
    \label{fig:krec_slice}
\end{figure*}

\citet{Biskamp2005a} derived the following criterion for conservation of magnetic topology in the context of resistive MHD:
\begin{equation}
    \label{eq:reconnection_criterion}
    \left\lVert\mathbf{B}\times\left(\nabla\times\mathbf{S}\right)\right\rVert = 0,
\end{equation}
where
\begin{equation}
    \label{eq:electric_field_projection}
    \mathbf{S} = \left(\frac{\mathbf{E} \cdot \mathbf{B}}{\mathbf{B} \cdot \mathbf{B}}\right)\mathbf{B}
\end{equation}
is the projection of the electric field onto the magnetic field direction. Reconnection takes place where Eq. \eqref{eq:reconnection_criterion} is violated. This can thus be used as a criterion for identifying reconnection sites. However, in the context of a numerical simulation, the onset of reconnection only occurs once the value
\begin{equation}
    \label{eq:krec}
    K = \left\lVert\mathbf{B}\times\left(\nabla\times\mathbf{S}\right)\right\rVert
\end{equation}
exceeds some finite threshold $K_\mathrm{min}$ due to limited precision in the employed numerical scheme. An example of how $K$ varies with position in our simulated atmosphere is shown in Fig. \ref{fig:krec_slice}.
We discuss how we determined a suitable value for $K_\mathrm{min}$ in Sect. \ref{sec:selecting_reconnection_sites}.

\subsubsection{Initial particle distributions}
\label{sec:initial_particle_distributions}

There is a range of possible mechanisms that can produce energetic particles during reconnection. One alternative is direct acceleration by the coherent electric field induced by the rotation of the magnetic field across the diffusion region \citep{Speiser1965, Martens1990, Litvinenko1993}. Test particle simulations of this kind of acceleration have been run for various magnetic configurations, including reconnecting Harris current sheets \citep{Zharkova2005a, Zharkova2005}, magnetic X-points \citep{Heerikhuisen2002, Wood2005}, and fan and spine reconnection \citep{Dalla2006, Dalla2008, Stanier2012}. These simulations generally produce particle populations with energy distributions that resemble power-laws.

Power-law distributions are also found in more realistic particle-in-cell simulations, which include the changes in the electric field induced by the accelerated particles in a self-consistent manner \citep{Baumann2013a, Li2019}. As can often be seen in these kinds of kinetic simulations, direct acceleration is likely to be accompanied by other types of acceleration processes. One example is first-order Fermi acceleration \citep{Fermi1954}, where particles gain energy by repeatedly scattering back and forth between converging magnetic elements, such as the ends of a shrinking plasmoid \citep{Drake2006} or in a collapsing magnetic trap \citep{Somov1997}. If the scattering agents move in a random rather than systematic fashion, second-order Fermi acceleration \citep{Fermi1949} can take place. Here, the particles experience a fluctuating energy increase owing to the higher likelihood of (accelerating) head-on collisions compared to (decelerating) rear-end collisions. This kind of stochastic acceleration is, like direct acceleration, typically predicted to produce a power-law energy distribution for the accelerated particles, both in models based on the Fokker--Planck formalism \citep{Miller1996, Stackhouse2018} and in test particle simulations \citep{Dmitruk2003, Onofri2006}.

It is widely accepted that energetic electrons play an important role in energy transport during solar flares. The detection of gamma-rays from large flares has revealed that also ions can play a part in these events \citep{Chupp1973}. However, for small flares, the effect of accelerated ions is likely to be minor. This is because ions, owing to their high masses, have significantly lower velocities than electrons for a given kinetic energy. As a consequence, they experience a much higher rate of collisions with ambient particles (the frequency of Coulomb collisions decreases with the cube of the velocity), and thus they lose their energy to the background plasma faster. For example, consider an electron or proton with mass $m$ travelling through an ionised hydrogen plasma with number density $n_\mathrm{H}$. The change in kinetic energy $E$ with distance $s$ for the particle is given by \citep[e.g.][]{Emslie1978}
\begin{equation}
    \label{eq:particle_energy_loss}
    \frac{\mathrm{d}E}{\mathrm{d}s} = -\left(\frac{m}{m_\mathrm{e}}\right)\frac{2\pi e^4 n_\mathrm{H}\ln\Lambda}{E},
\end{equation}
where $e$ is the elementary charge and $\ln\Lambda$ is the Coulomb logarithm (discussed further in Sect. \ref{sec:particle_energy_deposition}). It is clear from this that a proton ($m = m_\mathrm{p}$) deposits its energy much faster than an electron ($m = m_\mathrm{e}$), by a factor of $m_\mathrm{p}/m_\mathrm{e} \approx 1800$. Hence, unless the ions are accelerated to very high velocities, their energy can be expected to end up close to the reconnection sites. This suggests that it is safe to omit ions when modelling the long-range energy transport in small flares.

Simulating the formation of an accelerated electron distribution during a reconnection event is a computationally expensive task. When many events have to be considered, a detailed simulation of the acceleration is therefore not a viable option on current hardware. However, it is reasonable to assume that the result of the acceleration process will be a non-thermal population of electrons with energies distributed according to a power-law:
\begin{equation}
    \label{eq:non_thermal_distribution}
    n_\mathrm{NT}(E \geq E_\mathrm{c}) = n_\mathrm{acc}\left(\frac{\delta - 1/2}{E_\mathrm{c}}\right)\left(\frac{E}{E_\mathrm{c}}\right)^{-(\delta + 1/2)}.
\end{equation}
Here, $n_\mathrm{acc} = \int_{E_\mathrm{c}}^\infty n_\mathrm{NT}(E)\mathrm{d}E$ is the number density of accelerated electrons and $E_\mathrm{c}$ is a lower cut-off energy below which electrons are not considered non-thermal. The power-law index $\delta$ controls how rapidly the number of electrons diminishes with higher energy. It is usually defined in terms of the non-thermal electron flux $F_\mathrm{NT} = n_\mathrm{NT}v$ (where $v \propto E^{1/2}$ is the electron speed), so that $F_\mathrm{NT}(E) \propto E^{-\delta}$.

The range of possible values for the power-law index $\delta$ in Eq. \eqref{eq:non_thermal_distribution} is subject to some loose observational constraints. Spectral analysis of hard X-ray bursts has shown that the non-thermal bremsstrahlung emission due to the interactions of accelerated electrons with the ambient plasma tends to have a single or double power-law distribution in energy \citep[e.g.][]{Kane1970, Lin1987}. Working backwards from the observed spectrum one can attempt to infer the initial distribution of the non-thermal electrons by considering the bremsstrahlung emission process inside the X-ray source and the energy loss of the electrons during their journey to the source from the acceleration region \citep{Holman2011, Kontar2011}.

Studies of this type, both of regular flares \citep{Kontar2002, Kontar2003, Sui2005, Battaglia2006, Krucker2010} and microflares \citep{Lin2001, Krucker2002, Christe2008, Hannah2008, Glesener2020}, suggest that the initial distribution follows a power-law with $\delta$ varying between 2 and 10, typically with larger values for less energetic events. There is some observational evidence for a linear-log relationship between $\delta$ and the X-ray flux measured at a fixed energy \citep{Grigis2004}, which has been reproduced in numerical models of stochastic acceleration \citep{Grigis2005, Grigis2006}. However, in the absence of a proper acceleration simulation for predicting its value, the least speculative way of specifying $\delta$ is to treat it as a free parameter. Section \ref{sec:effect_p_delta} discusses the effect of varying $\delta$ in our model.

The total power $P_\mathrm{acc}$ going into the acceleration of an electron population generally corresponds to some fraction $p$ of the rate of magnetic energy release $P_\mathrm{rec}$ at the reconnection site:
\begin{equation}
    \label{eq:acceleration_power}
    P_\mathrm{acc} = p P_\mathrm{rec}.
\end{equation}
If the volume of the reconnection site is $V$, the average acceleration power per volume is
\begin{equation}
    \label{eq:acceleration_power_density}
    e_\mathrm{acc} = \frac{P_\mathrm{acc}}{V},
\end{equation}
and if the acceleration process lasts for a duration $\Delta t$, the energy density of accelerated electrons in the reconnection site is
\begin{equation}
    \label{eq:acceleration_energy_density}
    u_\mathrm{acc} = e_\mathrm{acc}\Delta t.
\end{equation}
This quantity is also related to the number density of non-thermal electrons:
\begin{equation}
    \label{eq:acceleration_energy_density_vs_number_density}
    u_\mathrm{acc} = \int_{E_\mathrm{c}}^\infty E\;n_\mathrm{NT}(E)\;\mathrm{d}E = n_\mathrm{acc}\left(\frac{2\delta - 1}{2\delta - 3}\right)E_\mathrm{c}.
\end{equation}
So knowledge of the fraction $p$ together with basic properties of the reconnection event enables the determination of $u_\mathrm{acc}$, which can be used to compute $n_\mathrm{acc}$ through Eq. \eqref{eq:acceleration_energy_density_vs_number_density}.

In a pure resistive MHD context, all the dissipated magnetic energy is released through Joule heating in the reconnection current sheets. Therefore, the local Joule heating prior to inclusion of particle acceleration can be used as a proxy for the reconnection energy, giving the relation
\begin{equation}
    \label{eq:acceleration_power_density_qjoule}
    e_\mathrm{acc} = p Q_\mathrm{Joule},
\end{equation}
where $Q_\mathrm{Joule}$ is the Joule heating rate per volume. Because a fraction $p$ of the energy that would previously go into Joule heating now is used for electron acceleration, $Q_\mathrm{Joule}$ must be reduced accordingly after application of Eq. \eqref{eq:acceleration_power_density_qjoule}.

Observational studies of the energy partition in flares suggest that typical values of $p$ could range from $10\%$ \citep{Emslie2004, Emslie2012} to as high as $50\%$ \citep{Lin1971}, and kinetic reconnection simulations support that values of these magnitudes indeed are conceivable \citep{Tsiklauri2007, Baumann2013a}. However, just like for $\delta$, the way $p$ depends on the details of the acceleration mechanism and the local conditions is subject to a great deal of uncertainty, so it is best kept as a free parameter. The effect of $p$ in our model is discussed in Sect. \ref{sec:effect_p_delta}.

The treatment of particle acceleration presented here builds on the premise that some unspecified acceleration mechanism will add a power-law tail with a known number density $n_\mathrm{acc}$ and index $\delta$ to the local thermal distribution of ambient electrons. This non-thermal component can then be isolated by defining the lower cut-off energy $E_\mathrm{c}$ as the energy where the power-law distribution intersects the thermal distribution. The original number density of thermal electrons should in principle be adjusted to account for some of them being accelerated. However, when dealing with the relatively minor energy releases associated with small flares, it is safe to assume that only a small fraction of the available electrons are accelerated. This correction can then be omitted.

\begin{figure}[!thb]
    \includegraphics{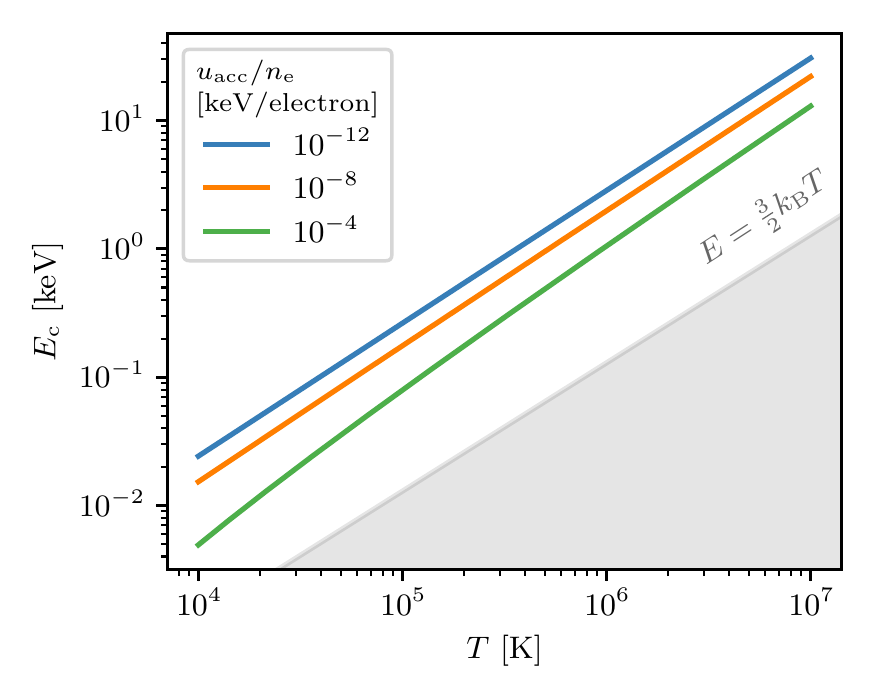}
    \centering
    \caption{Temperature dependence of the lower cut-off energy $E_\mathrm{c}$, for a selection of values of the non-thermal energy per thermal electron, $u_\mathrm{acc}/n_\mathrm{e}$, representative of the conditions in a relatively quiet atmosphere. A power-law index of $\delta = 4$ was used, but any realistic value would give practically identical results. The shaded area is where the cut-off energy would be lower than the average thermal energy.}
    \label{fig:Ec_parameter_study}
\end{figure}

The thermal electron population follows the Maxwell--Boltzmann distribution
\begin{equation}
    \label{eq:maxwell_boltzmann_distribution}
    n_\mathrm{T}(E) = n_\mathrm{e}\sqrt{\frac{4E}{\pi(k_\mathrm{B}T)^3}}e^{-E/k_\mathrm{B} T},
\end{equation}
where $n_\mathrm{e}$ is the number density of thermal electrons, $T$ is the local temperature, and $k_\mathrm{B}$ is the Boltzmann constant. The above definition of $E_\mathrm{c}$ can then be written as
\begin{equation}
    n_\mathrm{NT}(E_\mathrm{c}) = n_\mathrm{T}(E_\mathrm{c}).
\end{equation}
After inserting Eqs. \eqref{eq:non_thermal_distribution} and \eqref{eq:maxwell_boltzmann_distribution}, and substituting $n_\mathrm{acc}$ using Eq. \eqref{eq:acceleration_energy_density_vs_number_density} to remove the implicit dependence on $E_\mathrm{c}$, we find after some rearranging
\begin{equation}
    \label{eq:lower_cutoff_energy}
    {E_\mathrm{c}}^{5/2}e^{-E_\mathrm{c}/k_\mathrm{B} T} = (\delta - 3/2)\left(\frac{u_\mathrm{acc}}{n_\mathrm{e}}\right)\sqrt{\frac{\pi(k_\mathrm{B}T)^3}{4}}.
\end{equation}
This can be solved numerically for $E_\mathrm{c}$ using, for example, the Newton--Raphson method. Only the highest-energy solution is relevant in this case. The resulting cut-off energy is roughly proportional to temperature, but is not sensitive to $u_\mathrm{acc}$ or $n_\mathrm{e}$, as shown in Fig. (\ref{fig:Ec_parameter_study}). A temperature of $10^6$ K results in a cut-off energy of the order of 1 keV.

With the energy distribution of the non-thermal electrons in place, the next aspect to consider is their directions of motion. This can be described in terms of their distribution of pitch angles $\beta$, defined as the angle between the direction of motion $\hat{\mathbf{v}}$ and the magnetic field direction $\hat{\mathbf{B}}$:
\begin{equation}
    \cos\beta = \hat{\mathbf{v}}\cdot\hat{\mathbf{B}}.
\end{equation}
The pitch angle distribution, just like the energy distribution, depends on the nature of the acceleration mechanism. Typically, direct acceleration models predict that the non-thermal electrons have most of their velocity along the magnetic field direction, while stochastic acceleration models predict more isotropic populations.

When it comes to transport calculations, the simplest approach is to adopt the view of a peaked initial pitch angle distribution and assume that all the electrons accelerated at a given reconnection site will leave the site with the same initial magnitude of the pitch angle cosine $|\mu_0| = |\cos\beta_0|$. If the underlying acceleration mechanism is assumed to only affect the average speed $v_\parallel$ of the electrons parallel to the magnetic field, any deviation from $|\mu_0| = 1$ must come from the average perpendicular speed $v_\perp$ of the electrons before acceleration. This speed corresponds to the average thermal speed
\begin{equation}
    v_\perp = \sqrt{\frac{8 k_\mathrm{B} T}{\pi m_\mathrm{e}}}.
\end{equation}
The total average speed of the accelerated electrons can be written as
\begin{equation}
    v_\mathrm{mean} = \sqrt{{v_\perp}^2 + {v_\parallel}^2}.
\end{equation}
This speed can also be computed as the expected value of $v = \sqrt{2E/m_\mathrm{e}}$ for the power-law distribution, which becomes
\begin{equation}
    v_\mathrm{mean} = \frac{2\delta - 1}{2\delta - 2}\sqrt{\frac{2 E_\mathrm{c}}{m_\mathrm{e}}}.
\end{equation}
The average magnitude of the pitch angle cosine can then be estimated as
\begin{equation}
    |\mu_0| = \frac{v_\parallel}{v_\mathrm{mean}} = \sqrt{1 - \left(\frac{v_\perp}{v_\mathrm{mean}}\right)^2}.
\end{equation}
We note that the case $v_\perp = v_\mathrm{mean}$, and correspondingly, $\mu_0 = 0$, occurs for $E_\mathrm{c} \approx k_\mathrm{B} T$. As shown in Fig. \ref{fig:Ec_parameter_study}, $E_\mathrm{c}$ will usually exceed $k_\mathrm{B} T$ by about one order of magnitude, so this approach will tend to give $|\mu_0| \approx 1$ in practice.

The direction in which the electron beam leaves the acceleration region must also be determined. This can be parallel or anti-parallel to the magnetic field direction, or both, again depending on the nature of the acceleration mechanism. Without a more detailed specification of this mechanism, the method of deciding the directions will naturally be somewhat ad hoc. However, it seems reasonable that the overall electric field direction $\hat{\mathbf{E}}$ in the acceleration region could provide some indication. If $\hat{\mathbf{E}}\cdot\hat{\mathbf{B}}$ is close to $\pm 1$, one might expect most of the electrons to escape in the $\mp\mathbf{B}$-direction (the opposite sign is due to their negative charge). On the other hand, if $\hat{\mathbf{E}}\cdot\hat{\mathbf{B}}$ is closer to zero, there is no immediate reason to prefer one direction over the other, and the electrons would probably partition more evenly between both directions. Based on this, a sensible strategy is to split the available non-thermal power $P_\mathrm{acc}$ between a forward propagating beam ($+\hat{\mathbf{B}}$-direction) with power $P_\mathrm{beam}^+$ and a backward propagating beam with power $P_\mathrm{beam}^-$. The power can be partitioned in the following way:
\begin{equation}
    \label{eq:beam_power_partition}
    P_\mathrm{beam}^\pm = \frac{1 \mp \hat{\mathbf{E}}\cdot\hat{\mathbf{B}}}{2}P_\mathrm{acc}.
\end{equation}
So if, for example, $\hat{\mathbf{E}}\cdot\hat{\mathbf{B}} = -0.2$, the forward propagating beam gets $60\%$ of the power and the backward propagating beam gets $40\%$. At any reconnection site, $\hat{\mathbf{E}}\cdot\hat{\mathbf{B}}$ is necessarily non-zero, and the smallest possible magnitude it can have depends on the choice of $K_\mathrm{min}$.

\subsubsection{Particle energy deposition}
\label{sec:particle_energy_deposition}
A particle with charge $q$ leaving a reconnection site with velocity $\mathbf{v}$ experiences a Lorentz force $\mathbf{F} = q(\mathbf{E} + \mathbf{v} \times \mathbf{B})$ due to the local electric and magnetic field. The $\mathbf{v} \times \mathbf{B}$ term gives the particle a helical motion around the magnetic field direction, without affecting its kinetic energy. The relative magnitude of $\mathbf{v}$ and $\mathbf{B}$ decides the radius of the helical motion, which for a typical electron in a normal coronal environment is smaller than a metre.

If an electric field is present, the motion of the particle can be influenced in two different ways. Firstly, the particle will be accelerated along the magnetic field direction if the electric field component in this direction is non-zero. However, this can only take place in magnetic diffusion regions where ideal MHD breaks down. Secondly, the centre of the helical motion will drift away from the original field line if the electric field has a component perpendicular to the magnetic field. This effect is generally negligible, because the bulk plasma velocity $\mathbf{u}$ would have to be comparable to the particle velocity to induce an electric field $\mathbf{E} \approx -\mathbf{u} \times \mathbf{B}$ with a magnitude that is comparable to the $\mathbf{v} \times \mathbf{B}$ term.

When drift away from the field line is ignored, it is convenient to describe the particle's motion in terms of its kinetic energy $E$, pitch angle $\beta$, and one-dimensional position $s$ along the field line. Because the gyroradius of the particle generally is very small compared to its typical travel distance (which is of the order of megametres), the offset of the particle perpendicular to the field line can safely be disregarded. Additionally, the journey of the particle through the atmosphere is typically so brief that it can be considered instantaneous compared to the time scale of the atmosphere's response to the particle beam. For example, a beam of 1 keV electrons traverses a 10 Mm coronal loop in about 0.5 s, while a pressure change due to the heating at a footpoint would need $\sim 100$ s to propagate the same distance back along the loop (assuming a sound speed of $c_\mathrm{s} \approx 10^4\sqrt{T}\;\mathrm{cm}/\mathrm{s}$ and a temperature of $T = 10^6\;\mathrm{K}$).

When the particle enters a region with a stronger magnetic field, it starts to gyrate more rapidly around the field axis due to the increased $\mathbf{v} \times \mathbf{B}$ force. This force, being perpendicular to the direction of motion, does not affect the kinetic energy, so the velocity of the particle parallel to the field axis decreases accordingly. If the increase in the magnetic field strength becomes sufficiently large, the movement of the particle along the field will eventually stop and then continue in the opposite direction. This magnetic mirroring effect could thus potentially trap particles in the coronal part of a magnetic loop. However, since the coronal magnetic field strength typically increases relatively slowly with depth, magnetic trapping is unlikely to drastically inhibit the particles from reaching the lower atmosphere. Therefore, we have ignored the effect of varying magnetic field strength in this initial treatment of particle propagation.

As the particle travels along the field line it exchanges energy and momentum with the ambient plasma through Coulomb interactions, both with free electrons and ions, and with electrons bound in neutral atoms. Collectively, these types of collisions have the effect of reducing and randomising the velocities of the accelerated particles, until their distribution merges with the background thermal distribution. The energy loss of the particles manifests as a heating of the local plasma. A simple and widely used approach for modelling Coulomb collisions is to approximate the evolution of the energy and pitch angle of a single particle based on the mean rate of energy and pitch angle dissipation \citep[as derived by][]{Spitzer1962}. This was done by \citet{Brown1972} for non-thermal electrons in an ionised hydrogen plasma. \citet{Emslie1978} generalised Brown's treatment to allow for a hydrogen plasma with an arbitrary, but uniform, degree of ionisation, and also obtained the rate of energy deposition as a function of depth for the full population of accelerated electrons by convolving the mean energy loss of a single electron with the initial non-thermal number distribution. \citet{Hawley1994} showed how an approximation in the derivations of Emslie can allow for an ionisation degree that varies with depth without having to resort to numerical integration. Following their approach, the rate of energy deposition per volume, $Q$, at distance $s$ along the field line can be written as
\begin{equation}
    \label{eq:beam_heating_per_volume}
    Q(s) = n_\mathrm{H}(s)\left(\frac{\pi e^4 (\delta - 2) F_\mathrm{beam}}{|\mu_0| {E_\mathrm{c}}^2}\right)\gamma(s) B\left(\kappa(s); \frac{\delta}{2}, \frac{1}{3}\right)\left(\frac{N^*(s)}{N_\mathrm{c}^*}\right)^{-\delta/2}.
\end{equation}
This equation assumes that the electrons all have the same initial pitch angle cosine $\mu_0$ and initial energies given by a power-law distribution as described by Eq. \eqref{eq:non_thermal_distribution}. $F_\mathrm{beam}$ is the energy flux of the beam of accelerated electrons leaving the reconnection site. The quantity $\gamma$, given by
\begin{equation}
    \gamma(s) = x(s)\ln\Lambda + (1 - x(s))\ln\Lambda',
\end{equation}
is a hybrid Coulomb logarithm that merges the contribution of the free electron Coulomb logarithm $\ln\Lambda$ and the neutral hydrogen Coulomb logarithm $\ln\Lambda'$ depending on the local ionisation fraction $x(s)$. $B$ is the incomplete beta function, defined by
\begin{equation}
    B(\kappa; a, b) = \int_0^\kappa t^{a - 1}(1 - t)^{b - 1}\;\mathrm{d}t.
\end{equation}
The integration limit used for $B$ is a ramp function given by
\begin{equation}
    \kappa(s) = \mathrm{max}\left(\frac{N(s)}{N_\mathrm{c}(s)}, 1\right),
    \end{equation}
where
\begin{equation}
    N(s) = \int_0^s n_\mathrm{H}(s')\;\mathrm{d}s'
\end{equation}
is the hydrogen column depth and
\begin{equation}
    N_\mathrm{c}(s) = \frac{\mu_0 {E_\mathrm{c}}^2}{6\pi e^4 \gamma(s)}
\end{equation}
is the stopping column depth for an electron with energy $E_\mathrm{c}$. The ionised column depth $N^*(s)$ is defined analogously to $N(s)$ as
\begin{equation}
    N^*(s) = \int_0^s \left(\frac{\gamma(s')}{\ln\Lambda}\right)n_\mathrm{H}(s')\;\mathrm{d}s'.
\end{equation}
Similarly, the ionised stopping column depth $N_\mathrm{c}^*$ corresponds to $N_\mathrm{c}$ with $\gamma = \ln\Lambda$.

\begin{figure*}[!thb]
    \includegraphics{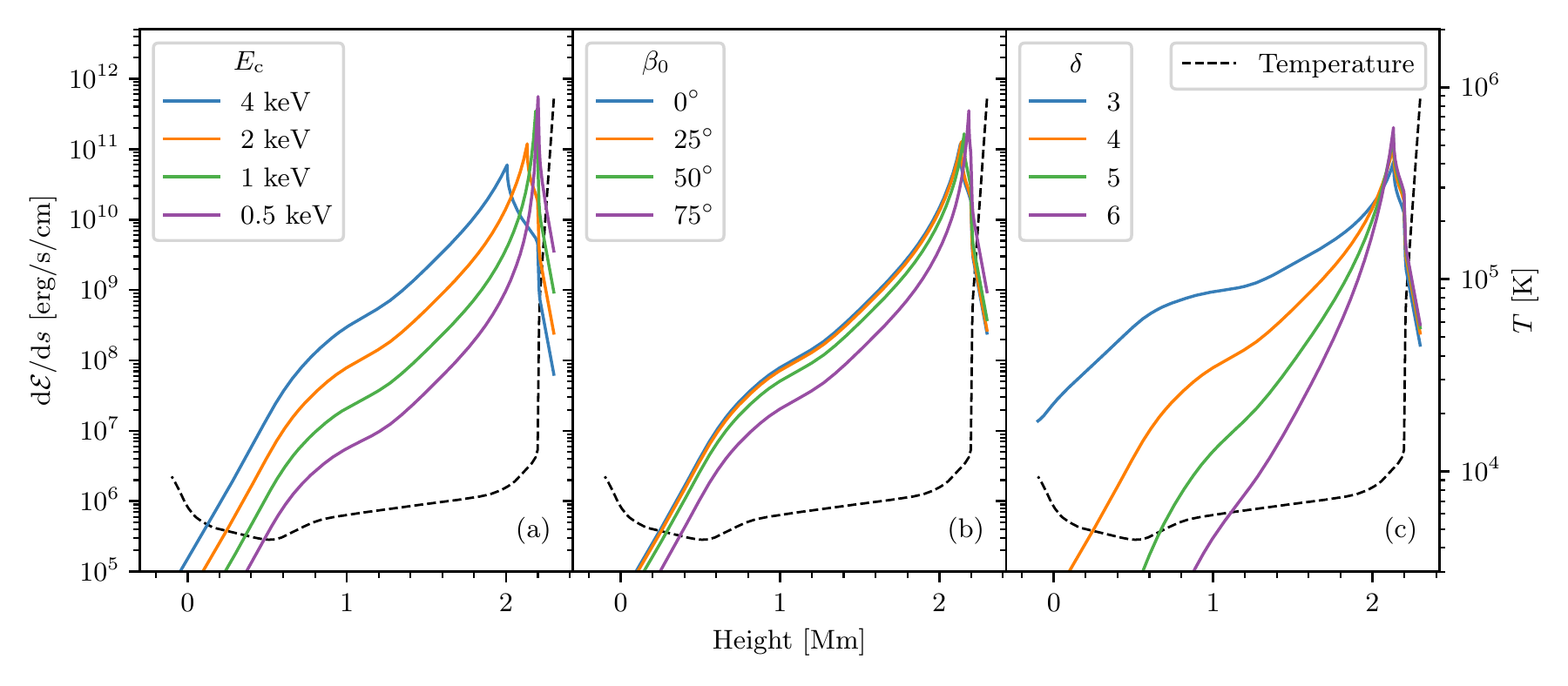}
    \centering
    \caption{Heating from an electron beam injected into the transition region of an average quiet sun atmosphere. The rate of energy deposition per distance is plotted against the height above the photosphere, for different values of the lower cut-off energy $E_\mathrm{c}$ (panel (a)), initial pitch angle $\beta_0$ (panel (b)) and power-law index $\delta$ (panel (c)). In all cases, the beam originates 2.3 Mm above the photosphere with a power of $P_\mathrm{beam} = 10^{18}\;\mathrm{erg}/\mathrm{s}$. Additionally we have used $E_\mathrm{c} = 2\;\mathrm{keV}$ for panels (b) and (c), $\beta_0 = 0^\circ$ for panels (a) and (c) and $\delta = 4$ for panels (a) and (b). The dashed curve is the temperature profile. The atmosphere corresponds to model C of \citet{Fontenla1993}, extrapolated to coronal temperatures.}
    \label{fig:single_beam_parameter_study}
\end{figure*}

The rate of energy deposition per distance, $\mathrm{d}\mathcal{E}/\mathrm{d}s$, can be found by integrating Eq. \eqref{eq:beam_heating_per_volume} over the cross-sectional area $A$ of the beam. If $Q(s)$ is assumed uniform across the beam cross-section, this gives $\mathrm{d}\mathcal{E}/\mathrm{d}s = AQ(s)$. Furthermore, if $A$ is assumed constant along the beam trajectory, it can be written as $A = P_\mathrm{beam}/F_\mathrm{beam}$, giving
\begin{equation}
    \label{eq:beam_heating_per_distance}
    \frac{\mathrm{d}\mathcal{E}}{\mathrm{d}s} = \left(\frac{P_\mathrm{beam}}{F_\mathrm{beam}}\right)Q(s).
\end{equation}
Fig. \ref{fig:single_beam_parameter_study} shows examples of the evolution of $\mathrm{d}\mathcal{E}/\mathrm{d}s$ with depth, computed from Eqs. \eqref{eq:beam_heating_per_volume} and \eqref{eq:beam_heating_per_distance}, for an electron beam injected into the FAL-C model atmosphere \citep{Fontenla1993}.

The Coulomb logarithm $\ln\Lambda$ emerges in the calculation of the mean rate of velocity change from Coulomb collisions between free particles, $\langle\mathrm{d}v/\mathrm{d}t\rangle$, which involves an integral of the differential collision cross-section over all impact parameters $b$ \citep[e.g.][]{Rosenbluth1957}. The long-range nature of the Coulomb force causes the integral to diverge in the limit of large $b$, but this can be resolved by considering the screening of the force at long distances due to the response of the nearby charge carriers to each particle's electrostatic field. This screening imposes a maximum value $b_\mathrm{max}$ on the impact parameter, enabling the integral for $\langle\mathrm{d}v/\mathrm{d}t\rangle$ to be solved. The solution is $\langle\mathrm{d}v/\mathrm{d}t\rangle \propto \ln\Lambda$, where $\Lambda = b_\mathrm{max}/b_\mathrm{min}$ and $b_\mathrm{min}$ is the minimum impact parameter. For a particle with charge $ze$ and speed $v$ interacting with a stationary particle with charge $Ze$, energy considerations give $b_\mathrm{min} = zZe^2/m v^2$, where $m$ is the reduced mass of the two particles. The Debye screening length $\lambda_\mathrm{D}$ is often used for $b_\mathrm{max}$. In the context of an energetic particle beam, a more appropriate choice might be the particle mean free path $\eta = v/\nu$, where $\nu = \sqrt{4\pi e^2 n_\mathrm{e}/m_\mathrm{e}}$ is the plasma frequency, or the gyroradius $r_\mathrm{g}$, depending on which is smallest \citep{Emslie1978}.

Although the Coulomb logarithm in principle varies with both particle energy, local conditions, and the masses of the colliding particles, the logarithmic scaling should keep these variations relatively small. Equation \eqref{eq:beam_heating_per_volume} consequently assumes $\ln\Lambda$ to be constant, so the value of $\ln\Lambda$ should simply be computed in each acceleration region and used throughout the transport calculations for the associated electron beam. Because the electrons will not experience very strong magnetic fields, one can expect that $\eta < r_\mathrm{g}$, and hence use $b_\mathrm{max} = \eta$. For collisions with ambient free electrons, we have $z = Z = 1$ and $m = m_\mathrm{e}/2$, so we get
\begin{equation}
    \label{eq:electron_coulomb_logarithm}
    \ln\Lambda = \ln\sqrt{\frac{{E_\mathrm{mean}}^3}{2\pi e^6 n_\mathrm{e}}},
\end{equation}
where we have used $v = \sqrt{2 E_\mathrm{mean}/m_\mathrm{e}}$, which is the speed corresponding to the mean energy
\begin{equation}
    E_\mathrm{mean} = \left(\frac{2\delta - 1}{2\delta - 3}\right)E_\mathrm{c}
\end{equation}
of the electrons in the initial distribution. Collisions with ambient protons only account for a tiny fraction of the electron velocity change $\langle\mathrm{d}v/\mathrm{d}t\rangle$ due to the high mass of protons compared to electrons, and are thus ignored in the derivations leading to Eq. \eqref{eq:beam_heating_per_volume}. For collisions with neutral hydrogen, the resulting energy loss rate can be expressed analogously to that of collisions with free electrons \citep[see e.g.][]{Mott1949a, Emslie1978}, with an effective Coulomb logarithm of
\begin{equation}
    \label{eq:neutral_hydrogen_coulomb_logarithm}
    \ln\Lambda' = \ln\left(\frac{2 E_\mathrm{mean}}{1.105 \chi}\right),
\end{equation}
where $\chi$ is the ionisation potential of hydrogen. For simplicity, the less important contributions from collisions with helium and heavier elements are not included here. The effective Coulomb logarithm for collisions with neutral helium is similar to Eq. \eqref{eq:neutral_hydrogen_coulomb_logarithm}, and has a comparable magnitude \citep{Evans1955a}. Considering the roughly 20\% abundance of helium, the inclusion of helium collisions would lead to at most a 20\% increase in the rate of energy deposition in the neutral regions of the atmosphere, and less in the partially ionised regions.

The treatment of Coulomb collisions outlined here disregards randomisation of energy and direction, which manifests as a diffusion of the energy and pitch angle distribution with propagation depth. As long as the speeds of the ambient particles are negligible compared to the speeds of the accelerated particles (the so-called cold-target approximation), energy and pitch angle diffusion are unimportant. This is because the target particles can be considered effectively stationary, leading to a deterministic evolution of each accelerated particle. \citet{Jeffrey2019} found that the cold-target approximation tends to underestimate the amount of energy deposited in the lower atmosphere compared to the results of a full warm-target model because the electrons that thermalise in the corona eventually will diffuse down to the lower atmosphere and deposit their energy there. However, this conclusion was based on work not including standard thermal conduction, and the inclusion of thermal conduction would mitigate some of the discrepancies between the cold- and warm-target models. Until the difference between these models has been investigated further, we do not implement the more computationally expensive warm-target treatment in our model. Moreover, when using the simple acceleration model presented in Sect. \ref{sec:initial_particle_distributions}, the lowest energies obtained for the accelerated electrons (which tend to come from sites with coronal temperatures) are typically of the order of 1 keV. This can be seen from Fig. \ref{fig:Ec_parameter_study}, which also shows that the target plasma would need to have a temperature of at least $10^7$ K in order for the average thermal energy to be comparable to the typical accelerated electron energy. Although the impact region over time could be heated to this temperature \citep[e.g.][]{Mariska1989, Allred2005}, the relatively low acceleration energies involved in minor flare events makes this unlikely.

A beam of energetic electrons departing from an acceleration region takes away negative charge and distributes it along its trajectory, leading to charge separation. This imbalance produces an electrostatic field that drives a counter-flowing return current of ambient electrons \citep{Knight1977, Emslie1980}. A steady state where the return current continuously compensates for the charge separation is reached on a timescale comparable to the electron--ion collision time \citep{Larosa1989}. Because the current associated with the beam then is cancelled by the return current, this mechanism prevents any induction of a significant electromagnetic field by the beam.

As long as the beam flux is weak, the energy loss incurred by the beam electrons from moving through the opposing electrostatic potential is negligible compared to their energy loss from Coulomb collisions with the ambient plasma. We confirmed this for our simulation by evaluating the energy loss contributions due to collisions and return currents (given respectively by Eqs. (4) and (6) in \citet{Emslie1980}) in the acceleration regions, where the return current energy loss is at its highest. The ratio of return current to collisional energy loss was found to be at most $10^{-4}$.

The accelerated electrons are also subject to a small radiative energy loss. They emit synchrotron radiation due to their gyrating motion around the magnetic field lines \citep{Petrosian1985} as well as bremsstrahlung due to collisions \citep{Brown1971, Haug2004}. A comparison between the energy loss terms from synchrotron and bremsstrahlung emission with the collisional loss term shows that both forms of radiative losses are completely negligible compared to collisional losses under ordinary conditions, and can safely be ignored.

There are a variety of considerations in addition to those covered above that a comprehensive particle transport model would need to address. This includes collisional ionisation of neutral chromospheric hydrogen \citep{Ricchiazzi1983, Abbett1999} and helium \citep{Allred2015}, the potential occurrence of a two-stream instability resulting in the generation of plasma oscillations and turbulence \citep{Emslie1984} as well as a fully relativistic treatment of the transport process \citep{McTiernan1990}. However, these effects tend to be more important for larger flares involving higher particle numbers and energies. For application to weaker acceleration events, the transport model presented here, in which only energy dissipation through Coulomb collisions is included, should be a reasonable first step.

\subsection{Model tuning}
\label{sec:tuning}

\subsubsection{Selection of reconnection sites}
\label{sec:selecting_reconnection_sites}
The method of identifying reconnection sites that is presented in Sect. \ref{sec:reconnection_sites} relies on an appropriate choice of the threshold $K_\mathrm{min}$. It should be set to a value small enough to include all the potentially important reconnection sites. However, it can not simply be set to zero, because limited spatial resolution and numerical diffusion in the MHD simulation prevent $K$ from ever becoming exactly zero in practice. Every point would then be classified as a reconnection site, which would be both unrealistic and prohibitively computationally expensive.

\begin{figure}[!thb]
    \includegraphics{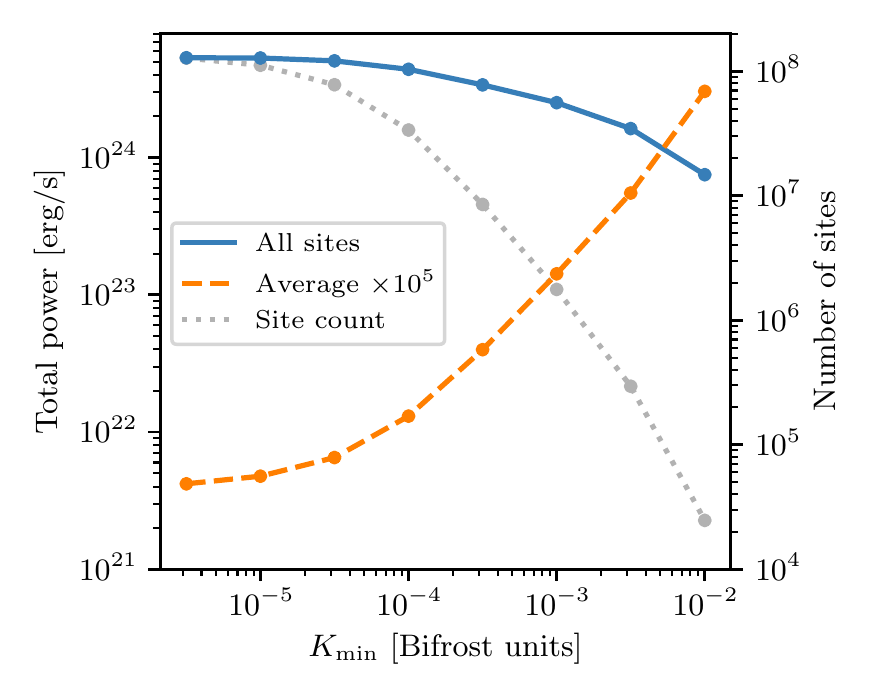}
    \centering
    \caption{Variation in acceleration power (Eq. \eqref{eq:acceleration_power}) with the reconnection factor threshold $K_\mathrm{min}$, for a simulation with $\delta = 4$ and $p = 0.2$. The solid curve is the total power for all included reconnection sites, and the dashed curve is the average power per site (multiplied by $10^5$ for composition purposes). The total numbers of included sites are shown in the dotted curve. Short-range beams have been filtered out in the manner discussed in Sect. \ref{sec:short_range_exclusion}.}
    \label{fig:global_heating_krec_lim}
\end{figure}

From Eqs. \eqref{eq:electric_field_projection} and \eqref{eq:krec} it can be seen that $K$ scales linearly with the strength of the magnetic field, $B$. The magnetic energy density $u_\mathrm{B}$ is proportional to $B^2$, meaning that $K$ is proportional to $\sqrt{u_\mathrm{B}}$. As $K_\mathrm{min}$ is lowered, the additional reconnection sites that are included thus produce less energetic particle distributions on average. On the other hand, the number of included sites also increases rapidly with decreasing $K_\mathrm{min}$. The choice of $K_\mathrm{min}$ is thus a compromise between the inclusion of more reconnection energy and the computational cost of simulating more electron beams. Fortunately, as shown in Fig. \ref{fig:global_heating_krec_lim}, the growth in the number of sites is balanced by the decrease in energy, and the total energy contained in all included beams begins to stagnate as $K_\mathrm{min}$ becomes sufficiently small. As a reasonable trade-off, we used $K_\mathrm{min} = 10^{-4}$ (in internal Bifrost units) for our results.

\subsubsection{Exclusion of short-range beams}
\label{sec:short_range_exclusion}
Not all of the identified acceleration regions produce electron beams that are worth considering. Most importantly, beams that deposit all their energy in the immediate vicinity of the acceleration region add nothing to the model. This is because the slight displacement of heat quickly would be evened out by other energy transport mechanisms such as thermal conduction, plasma advection, or radiative transfer. The outcome would thus be nearly the same as if all the reconnection energy had been converted directly into thermal energy at the reconnection site in the first place.

\begin{figure}[!thb]
    \includegraphics{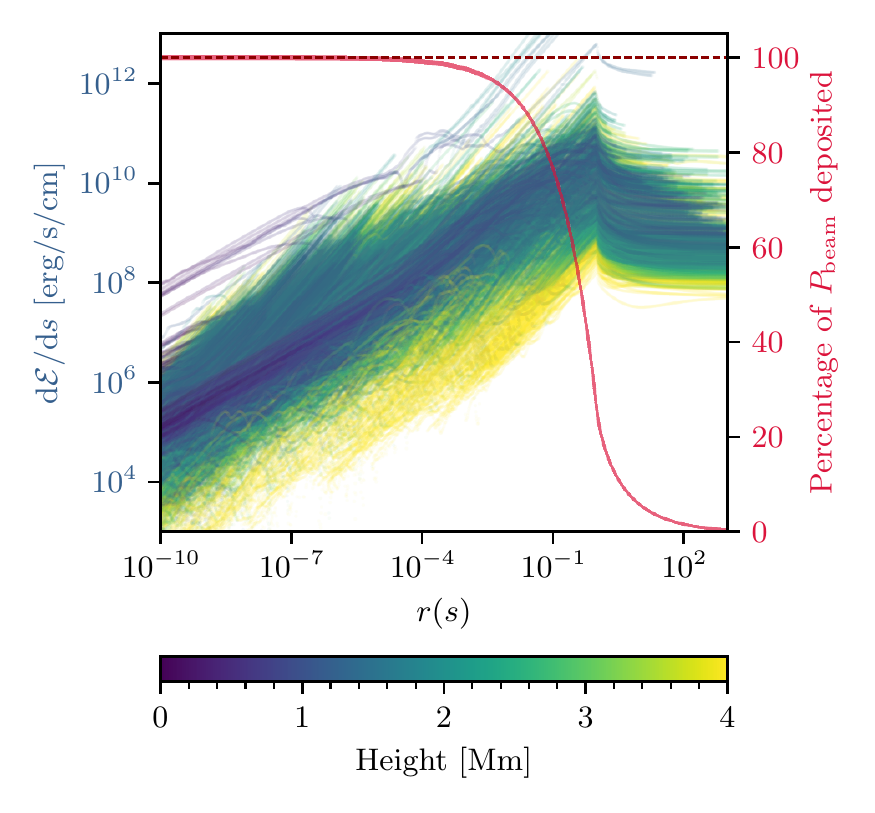}
    \centering
    \caption{Energy deposition as a function of $r(s)$ (Eq. \eqref{eq:residual_factor}). The deposited power per distance, $\mathrm{d}\mathcal{E}/\mathrm{d}s$, is plotted along the trajectories of a representative subset of the electron beams in a simulation with $\delta = 4$ and $p = 0.2$, with colours indicating the height above the photosphere. For each beam, the corresponding proportion of the initial beam power that has been deposited at each $r(s)$, given by ${P_\mathrm{beam}}^{-1}\int_0^s \mathrm{d}\mathcal{E}/\mathrm{d}s(s')\;\mathrm{d}s'$, is indicated by a red curve. These red curves are all overlapping.}
    \label{fig:deposited_percentage_vs_residual_factor}
\end{figure}

To filter out the short-range beams, we first had to establish a criterion for when a beam is considered depleted. It can be seen from Eq. \eqref{eq:beam_heating_per_volume} that $Q(s)$ approaches zero only asymptotically with distance. Physically, this can be explained by the presence of arbitrarily energetic electrons in the tail of the power-law distribution. Because the collisional cross-section decreases with electron energy, extremely energetic electrons will practically never thermalise, and hence there will always be some non-thermal energy remaining in the beam. However, once $Q$ becomes sufficiently small, the rest of the beam energy can safely be disregarded, provided that the reason for the small heating rate is the depletion of energy and not that the beam happens to pass through a low-density region. This second criterion can be ensured by considering the part of Eq. \eqref{eq:beam_heating_per_volume} representing energy depletion, which is the monotonically decreasing factor
\begin{equation}
    \label{eq:residual_factor}
    r(s) = \left(\frac{N_*(s)}{N_\mathrm{c}^*}\right)^{-\delta/2}.
\end{equation}
This is a convenient heuristic for the amount of energy remaining in the beam, as shown in Fig. \ref{fig:deposited_percentage_vs_residual_factor}, where the percentage of the initial beam power that has been deposited can be seen to approach $100\%$ as $r$ becomes smaller.
Using $(\mathrm{d}\mathcal{E}/\mathrm{d}s)_\mathrm{min}$ and $r_\mathrm{min}$ to denote lower thresholds for $\mathrm{d}\mathcal{E}/\mathrm{d}s$ and $r$, respectively, a depletion criterion can thus be defined as
\begin{equation}
    \label{eq:depletion_criterion}
    \frac{\mathrm{d}\mathcal{E}}{\mathrm{d}s}(s) < \left(\frac{\mathrm{d}\mathcal{E}}{\mathrm{d}s}\right)_\mathrm{min}\quad \mathrm{and} \quad r(s) < r_\mathrm{min}.
\end{equation}
It is clear from the figure that the vast majority of the initial beam power is depleted once $r$ is below $\sim 10^{-5}$. Therefore, this paper uses $r_\mathrm{min} = 10^{-5}$. Moreover, we set $(\mathrm{d}\mathcal{E}/\mathrm{d}s)_\mathrm{min} = 10^5\;\mathrm{erg}/\mathrm{s}/\mathrm{cm}$. As the figure shows, this enables the beams to reach deep into the lower atmosphere before they are considered depleted.

Based on the criteria in Eq. \eqref{eq:depletion_criterion}, an estimate $\tilde{s}_\mathrm{dep}$ for the depletion distance can be computed under the assumption that the plasma properties are approximately uniform between $s = 0$ and $s = s_\mathrm{dep}$, so that $N^*(s_\mathrm{dep}) \approx (n_\mathrm{H}(s=0)\gamma(s=0)/\ln\Lambda) s_\mathrm{dep}$. This assumption holds as long as $s_\mathrm{dep}$ is reasonably short. Equations \eqref{eq:beam_heating_per_volume}, \eqref{eq:beam_heating_per_distance}, \eqref{eq:residual_factor}, and \eqref{eq:depletion_criterion} then yield the following estimate for the depletion distance:
\begin{equation}
    \label{eq:estimated_depletion_distance}
    \tilde{s}_\mathrm{dep} = \left(\frac{N_\mathrm{c}^*\ln\Lambda}{n_\mathrm{H}(0)\gamma(0)}\right)\mathrm{max}\left(c_Q, c_r \right)^{2/\delta},
\end{equation}
where
\begin{equation}
    c_Q = \frac{n_\mathrm{H}(0) \gamma(0)}{(\mathrm{d}\mathcal{E}/\mathrm{d}s)_\mathrm{min}}\left(\frac{\pi e^4 (\delta - 2) P_\mathrm{beam}}{|\mu_0| {E_\mathrm{c}}^2}\right) B\left(1; \frac{\delta}{2}, \frac{1}{3}\right)
\end{equation}
and
\begin{equation}
    c_r = \frac{1}{r_\mathrm{min}}.
\end{equation}
The derivation also assumes that $\kappa(s_\mathrm{dep}) = 1$, which is always satisfied when $r < 1$. By evaluating Eq. \eqref{eq:estimated_depletion_distance} at each reconnection site, it can be decided whether the resulting electron beam is worth considering further. The beam can be excluded if $\tilde{s}_\mathrm{dep}$ is shorter than an assigned minimum distance $s_\mathrm{min}$.
\begin{figure}[!thb]
    \includegraphics{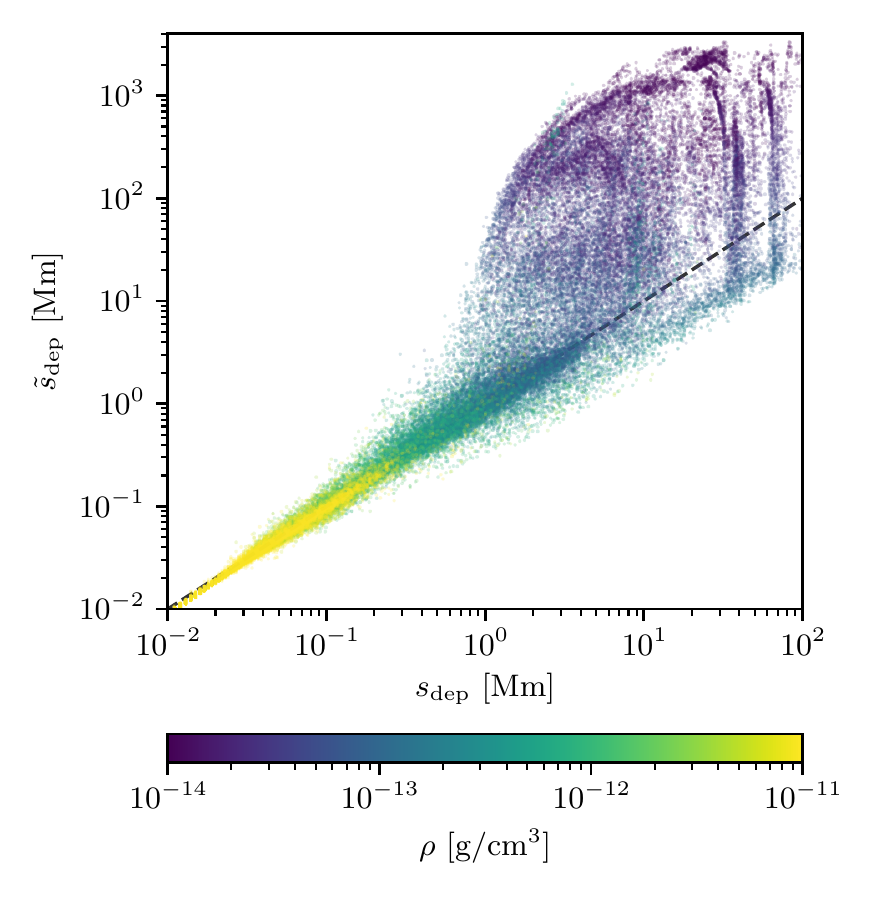}
    \centering
    \caption{Estimated depletion distances $\tilde{s}_\mathrm{dep}$ plotted against actual depletion distances $s_\mathrm{dep}$ for a representative subset of the electron beams in the same simulation as in Fig. \ref{fig:deposited_percentage_vs_residual_factor}. Points lying on the dashed line correspond to correct estimates. The colour indicates the mass density $\rho$ in the acceleration region.}
    \label{fig:depletion_distances}
\end{figure}
Figure \ref{fig:depletion_distances} confirms that Eq. \eqref{eq:estimated_depletion_distance} is accurate for small values of $s_\mathrm{dep}$. In the cases when $s_\mathrm{dep}$ is over-estimated, it could lead to the inclusion of a beam that turns out to propagate shorter than expected, but this has no impact on the accuracy of the result. More problematically, an under-estimation of $s_\mathrm{dep}$ could lead to the rejection of a beam that indeed would contribute to the long-range energy transport. However, cases like these appear to be relatively uncommon. With Fig. \ref{fig:depletion_distances} as a guideline, we chose a minimum distance of $s_\mathrm{min} = 0.5\;\mathrm{Mm}$ for our results. We note that the figure shows a clear inverse relationship between depletion distance and density, and that practically all beams accelerated at densities higher than about $10^{-12}\;\mathrm{g}/\mathrm{cm}^3$ will be rejected. Consequently, all non-thermal energy that is transported a significant distance in this model comes from the corona and upper transition region.

\subsubsection{Exclusion of low-energy beams}
\label{sec:weak_exclusion}
\begin{figure}[!thb]
    \includegraphics{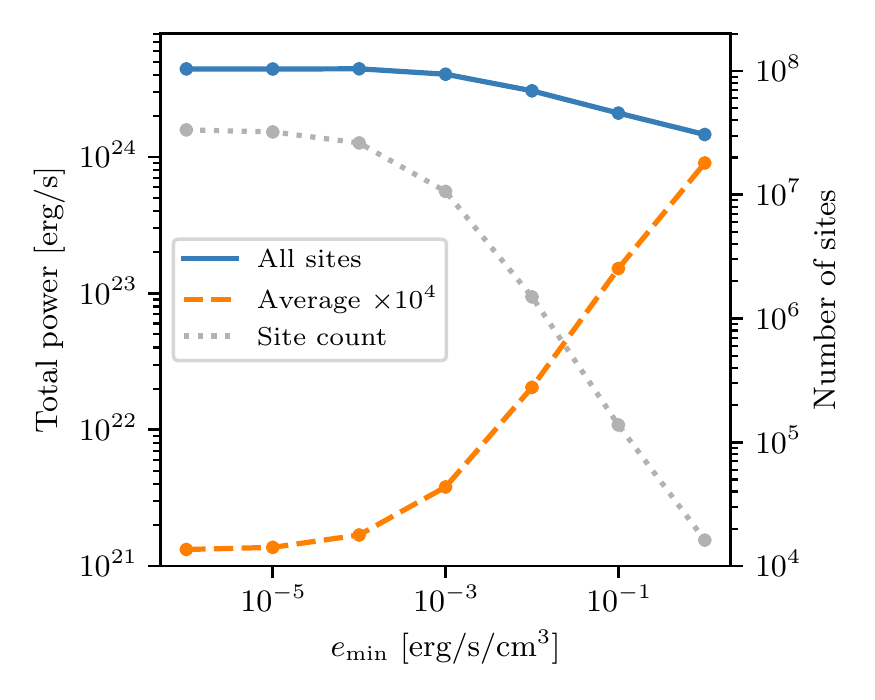}
    \centering
    \caption{Variation in global acceleration power with the lower limit $e_\mathrm{min}$ on the local acceleration power density. Like in Fig. \ref{fig:global_heating_krec_lim}, the simulated beams have $\delta = 4$ and $p = 0.2$, and the method discussed in Sect. \ref{sec:short_range_exclusion} was used to filter out short-range beams. The curves have the same meaning as in Fig. \ref{fig:global_heating_krec_lim}.}
    \label{fig:global_heating_min_beam_en}
\end{figure}

As discussed in Sect. \ref{sec:selecting_reconnection_sites}, the reconnection factor $K$ correlates with the available magnetic energy. However, because $K$ also depends on the configuration of the electromagnetic field, there will still be reconnection sites with $K > K_\mathrm{min}$ that have very low acceleration energies. As a way of reducing computational cost, these sites can be excluded with little consequence for the accuracy of the model by imposing a suitable lower limit $e_\mathrm{min}$ on the acceleration power density $e_\mathrm{acc}$ in Eq. \eqref{eq:acceleration_power_density_qjoule}. The total power in all included acceleration regions saturates as $e_\mathrm{min}$ is reduced, as shown in Fig. \ref{fig:global_heating_min_beam_en}.
Much like the situation in Fig. \ref{fig:global_heating_krec_lim}, this is because the increase in the number of included sites is counterbalanced by the decrease in the average power at each site. For very small values of $e_\mathrm{min}$, no additional beams are excluded, so the total power reaches a constant value. For this paper, we used $e_\mathrm{min} = 10^{-2}\;\mathrm{erg}/\mathrm{s}/\mathrm{cm}^3$, which lead to a drastic reduction in the number of acceleration regions with a relatively minor loss in included power.

\subsection{Effect of $p$ and $\delta$}
\label{sec:effect_p_delta}
Variation in the acceleration power fraction $p$ in Eq. \eqref{eq:acceleration_power} effectively leads to a proportional scaling of the energy deposition $Q(s)$ at every depth.\footnote{In principle, the relation is not directly proportional since $p$ also affects the cut-off energy $E_\mathrm{c}$ through $u_\mathrm{acc}$ in Eq. \eqref{eq:lower_cutoff_energy}. However, as Fig. \ref{fig:Ec_parameter_study} shows, this dependence is so weak as to be negligible.} Therefore, $p$ does not affect the spatial distribution of deposited beam energy, and the exact choice of its value has a limited qualitative bearing on the energy transport. Of course, if $p$ was extremely small, any effect of non-thermal electrons would be completely negligible regardless of how the electrons distributed their energy. Yet, based on the current understanding of the acceleration mechanisms taking place during reconnection, this seems unlikely. A value of $p = 0.2$ was therefore chosen for this paper.

In contrast to $p$, the choice of $\delta$ has a major influence on the resulting spatial distribution of deposited beam energy. Panel (c) in Fig. \ref{fig:single_beam_parameter_study} shows that a larger value of $\delta$ leads to a significantly faster rate of energy deposition with distance, and thus to a shorter penetration depth for the beam. This can be understood mathematically from the $-\delta/2$ power in Eq. \eqref{eq:beam_heating_per_volume}, and physically from the lower fraction of electrons in the high-energy tail of the non-thermal distribution. Because $\delta$ has the unfortunate feature of being both important and uncertain, we present results for a range of $\delta$-values where appropriate. Otherwise, we used a value of $\delta = 4$, which aids the analysis of the energy transport by giving the beams some penetrative power, while still being a realistic value lying well within the observed range.

\section{Results}
\label{sec:results}

\subsection{Global energy transport}
\label{sec:global_transport}
\begin{figure*}[!thb]
    \includegraphics{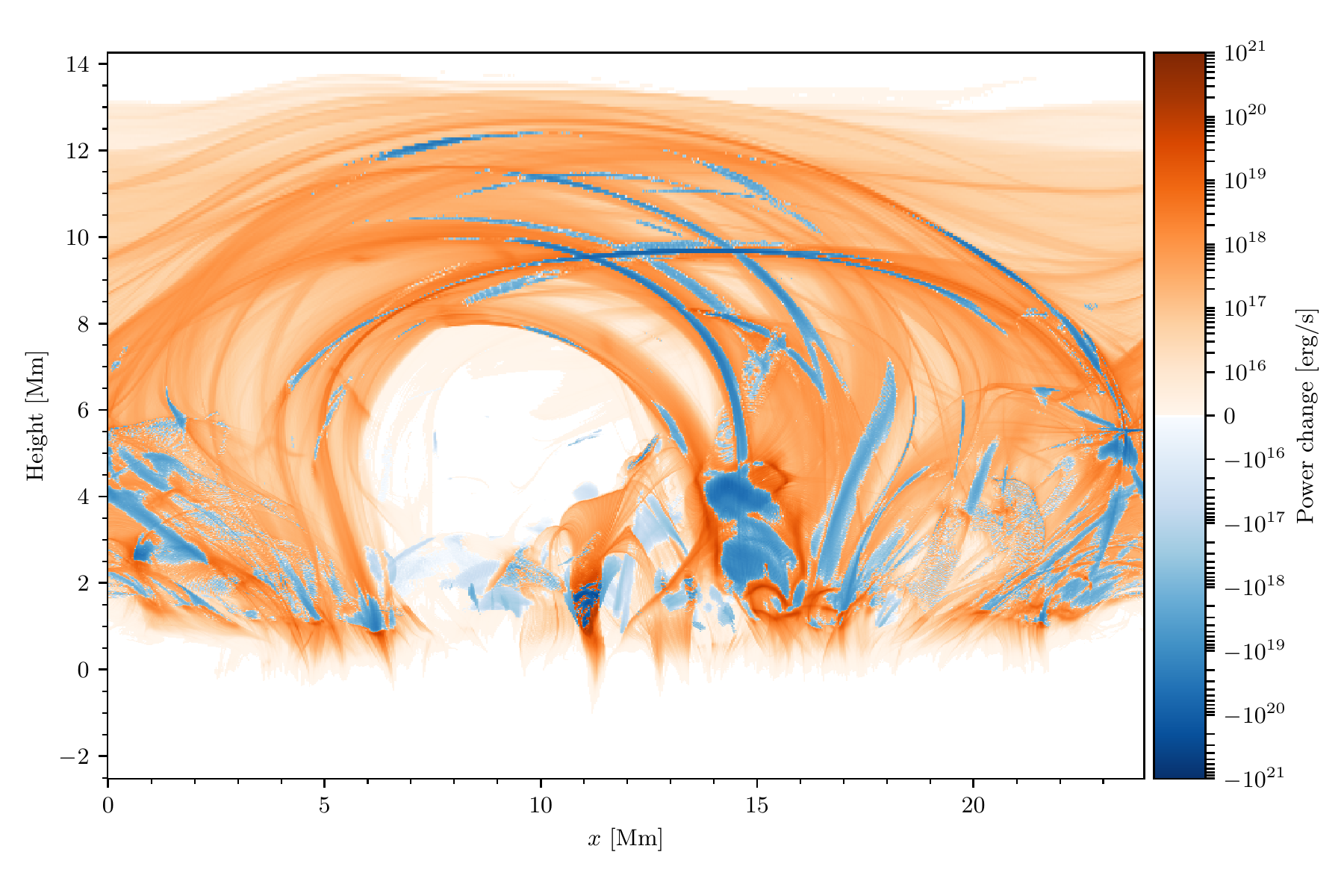}
    \centering
    \caption{Change in heating power in each grid cell due to the inclusion of acceleration and transport of non-thermal electrons. The power changes are accumulated along the $y$-axis of the simulation snapshot. Blue regions indicate a net reduction of thermal energy compared to the case without non-thermal electrons, which is due to the fraction $p$ of the local reconnection energy being injected into accelerated electrons instead of heating the ambient plasma. Orange regions show where the non-thermal electron energy eventually is deposited into the plasma as heat. The simulated beams have $\delta = 4$ and $p = 0.2$.}
    \label{fig:xz_power_change_beams}
\end{figure*}

\begin{figure*}[!thb]
    \includegraphics{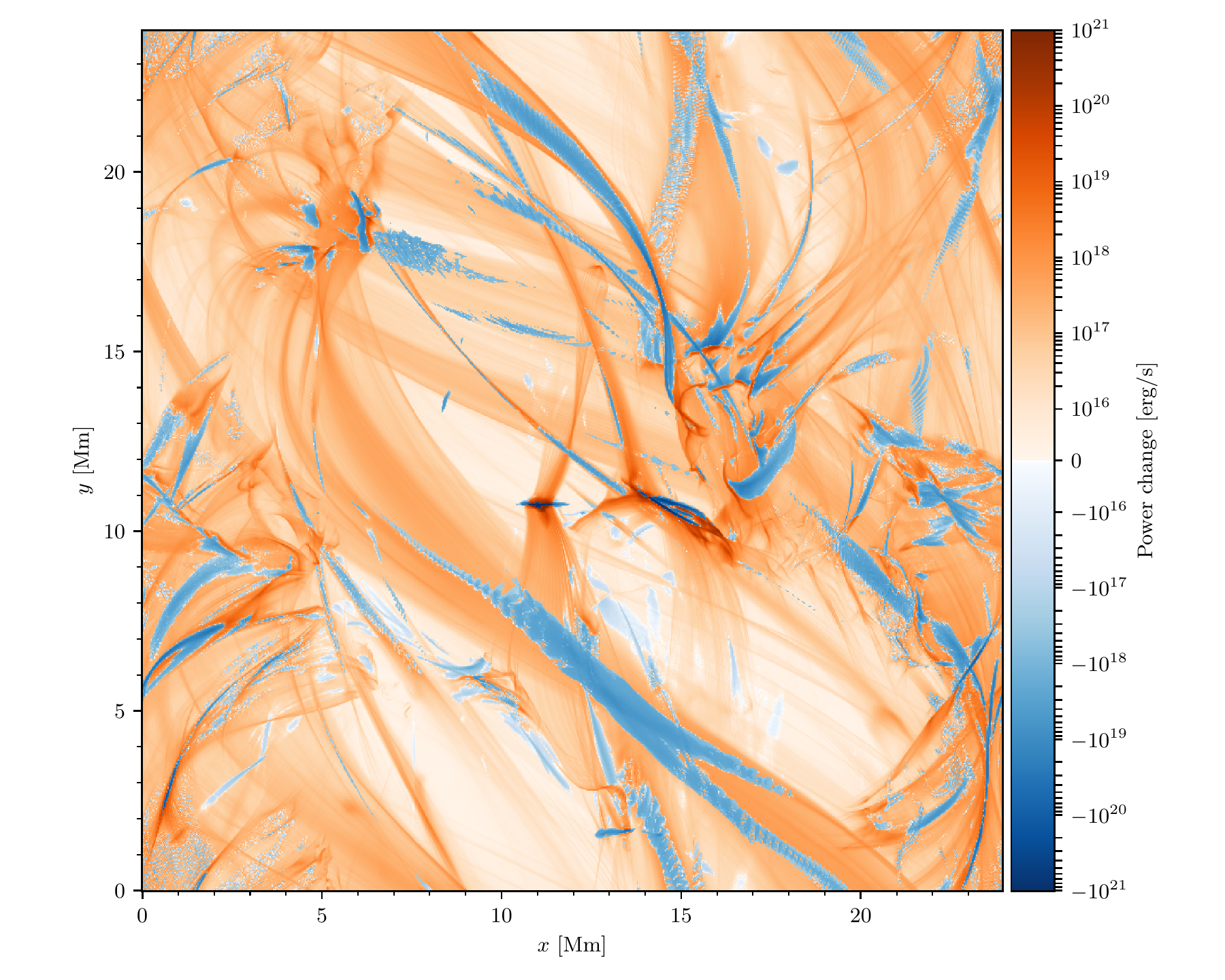}
    \centering
    \caption{Same as Fig. \ref{fig:xz_power_change_beams}, but with the power changes accumulated over the full height of the simulation snapshot instead of the $y$-axis.}
    \label{fig:horizontal_power_change_beams}
\end{figure*}

Particle acceleration predominantly occurs in localised regions that are aligned with the major magnetic field structures. This can be seen in Figs. \ref{fig:xz_power_change_beams} and \ref{fig:horizontal_power_change_beams}, which show the net electron beam heating power accumulated respectively horizontally and vertically over the simulation domain.
Due to the exclusion of short-range electron beams discussed in Sect. \ref{sec:short_range_exclusion}, all significant acceleration takes place within the tenuous plasma above the transition region. The acceleration regions (apparent as blue areas in the figures) have lengths ranging from 1 to 15 Mm, and cross-sections typically smaller than 1 Mm. Longer acceleration regions tend to occur higher in the corona, where the magnetic field is more homogeneous. Interestingly, despite the decrease of magnetic field strength with height (panel (c) in Fig. \ref{fig:atmospheric_height_profiles}), these high regions typically exhibit equally energetic acceleration as regions at lower heights. The most intense acceleration can be found in a thin sheet centred on $x = 11$ Mm and $y = 10.67$ Mm (the $y$-coordinate is the same as for the plane of Fig. \ref{fig:krec_slice}). This is the current sheet associated with the Ellerman bomb and UV burst that are analysed by \citet{Hansteen2019}.

Energy deposition from the non-thermal electrons (shown in orange in Figs. \ref{fig:xz_power_change_beams} and \ref{fig:horizontal_power_change_beams}) takes place throughout the corona, with higher concentrations near the ends of the acceleration regions and along the dominant magnetic structures. The strongest non-thermal heating typically occurs in the transition region near low-lying acceleration regions. At these locations, the electrons are often able to reach significant chromospheric depths. Numerous electron beams entering the lower atmosphere from different directions aggregate horizontally due to the convergence of the magnetic field with depth. This produces collections of thin, semi-vertical strands of concentrated non-thermal heating in the chromosphere, which are anchored in the photosphere at locations with a strong vertical magnetic field.

\subsection{Selected sets of electron beams}
\label{sec:selected_beams}
\begin{figure}[!thb]
    \centering
    \includegraphics{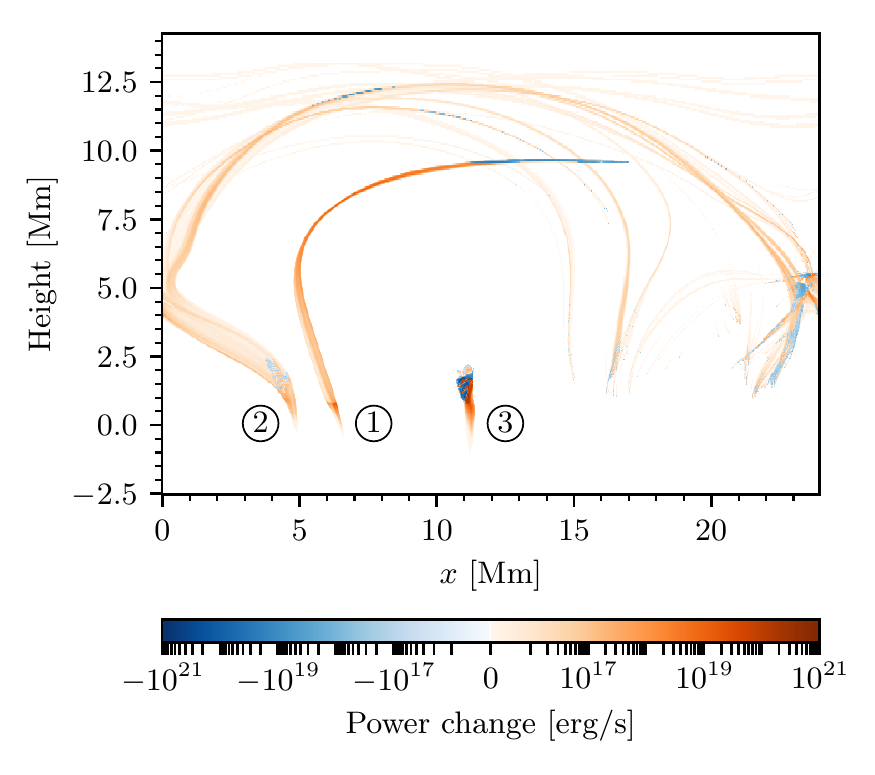}
    \caption{Selected sets of electron beams, plotted in the same manner as Fig. \ref{fig:xz_power_change_beams}. Set 1 is a coherent bundle of beams that originates in a long acceleration region at the top of a coronal loop and terminates at one of the footpoints. Set 2 consists of several electron beam bundles coming from various locations in the simulation domain, all converging at the same chromospheric site. Set 3 encompasses electrons that are accelerated in the strong central current sheet and ejected along one of the magnetic 'legs' that connect the current sheet with the lower chromospheric plasma.}
    \label{fig:xz_power_change_selected_beams}
\end{figure}

To analyse the beam heating in the lower atmosphere more closely, we consider the three subsets of electron beams shown in Fig. \ref{fig:xz_power_change_selected_beams}.
They represent various ways in which electron beams can join together to produce significant localised heating in the lower atmosphere. This includes a single long bundle originating high up in the corona (set 1), the convergence of multiple thin bundles coming from separate acceleration regions (set 2), and a short bundle associated with an acceleration region that lies just above the transition region (set 3).

\begin{figure*}[!thb]
    \includegraphics{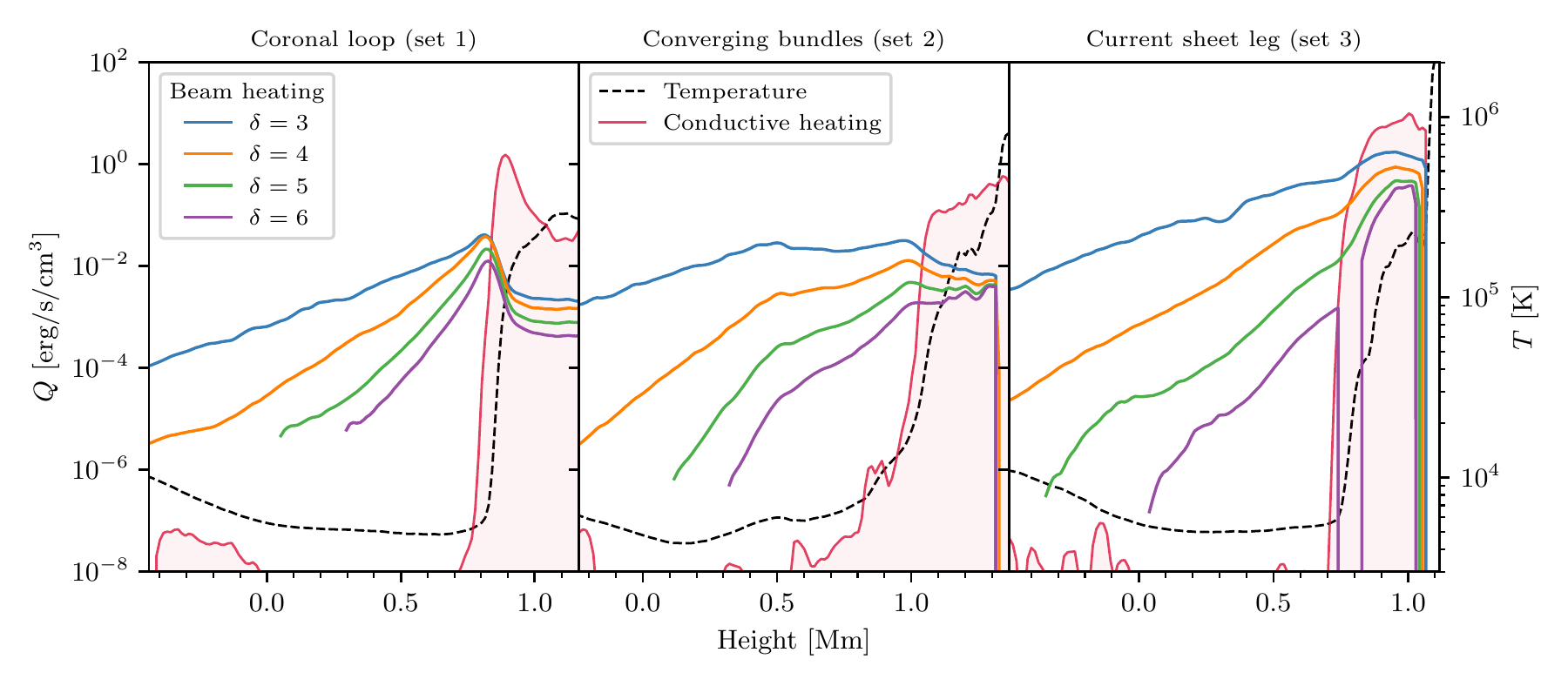}
    \centering
    \caption{Horizontal averages of heating due to electron beams and thermal conduction with height along the three selected sets of electron trajectories. The averages are taken over the grid cells for which the aggregated $Q_\mathrm{beam}$ exceeds its 75th percentile within each horizontal layer. The dashed line shows the corresponding average temperature profile. We note that the height axes have the same extent, but separate bounds.}
    \label{fig:heating_comparison}
\end{figure*}

In Fig. \ref{fig:heating_comparison}, the horizontal average of $Q_\mathrm{beam}$ in the core of the cone that penetrates the lower atmosphere is plotted with height for each beam set. In order to demonstrate the effect of the power-law index $\delta$, each beam heating profile is plotted for $\delta$ ranging from 3 to 6. The shapes of the local transition regions are apparent from the included temperature profiles.

Transition region beam heating can be seen to be relatively robust to variations in $\delta$. At the same time, this is where the difference between the origins of the electron beams for the three beam sets primarily manifests. The reason for this is that the number of incoming low- and intermediate-energy electrons, which make up the bulk of the transition region heating, does not change considerably with $\delta$, but is highly sensitive to the amount of coronal plasma that the beam has propagated through.

The electrons accelerated at the top of the coronal loop (set 1) produce a pronounced peak in the beam heating, centred on the bottom of the transition region. This is because most electrons with too little energy to make it through the transition region already would have stopped on the way through the coronal loop. For the converging electron beam bundles coming from separate locations (set 2), the corresponding peak is less distinct. Some of the beams in this set stem from just above the transition region, and their low-energy electrons provide a significant amount of heat to the upper transition region, making  the peak less pronounced. This situation is most evident for the beams coming from the strong current sheet that resides in the lower corona near the centre of the simulation domain (set 3). Here, the full spectrum of electron energies is injected directly into the transition region. As a result, the heating culminates near the top of the transition region and decreases monotonically with depth in the lower atmosphere.

Below the transition region, the decrease with depth of the average beam heating rate is highly dependent on $\delta$. At a given chromospheric depth, the decrease in $Q_\mathrm{beam}$ with increasing $\delta$ appears to be approximately exponential. However, the average slopes of the beam heating profiles for a given value of $\delta$ are similar for all three beam sets. They are slightly steeper for sets 1 and 3 than for set 2, but this is because the mass densities at these locations are somewhat higher.

In contrast to the situation in the transition region, only electrons in the high-energy tail of the incoming distribution can significantly penetrate the chromosphere. This explains the sensitivity to $\delta$, which controls the relative portion of high-energy electrons in the distribution. The electrons in the high-energy tail are not significantly influenced by the coronal plasma, so the distance travelled by the electrons through the corona has little bearing on the distribution of electrons that enter the chromosphere. As a result, any difference between the shapes of the chromospheric heating profiles, barring local variations in mass density, must be caused by a difference between the shapes of the initial electron distributions in the acceleration regions. In the acceleration model used here, this requires significant variations between the temperatures of the involved acceleration regions, which would lead to different values of the lower cut-off energy $E_\mathrm{c}$ (Fig. \ref{fig:Ec_parameter_study}). The selected beam sets all have similar temperatures in their acceleration regions, so the shape of the electron distribution that penetrates the chromosphere is comparable in all three cases.

Energy transport by accelerated electrons plays a similar role as thermal conduction, in that it transports energy from the corona to the transition region along the magnetic field. However, the relative importance of these mechanisms differs greatly with depth in the lower atmosphere. This is evident from the red curve in Fig. \ref{fig:heating_comparison}, which shows the conductive heating along the three sets of beam trajectories. In all cases, conductive heating is 10--100 times stronger than electron beam heating throughout most of the transition region. The strong conductive heating stems from the abrupt drop from coronal to chromospheric temperatures. But as the temperature decreases towards the chromosphere, so does the thermal conductivity of the plasma, causing the conductive heating to nearly vanish at the bottom of the transition region. On the other hand, the heat deposited by non-thermal electrons is close to its peak value at this location, owing to the sudden rise in mass density with depth. For all values of $\delta$, beam heating exceeds conductive heating by many orders of magnitude throughout the chromosphere.

\subsection{Energetics}
The total power of all accelerated electrons in the simulation snapshot is roughly $10^{24}\;\mathrm{erg}/\mathrm{s}$. Beam sets 1 and 2 each produce approximately $3\cdot 10^{21}\;\mathrm{erg}/\mathrm{s}$ of non-thermal electron power. This value is representative of a typical collection of electron beams forming a coherent lower atmospheric heating site in the simulation. Beam set 3, which is associated with a particularly energetic event, exceeds this power by two orders of magnitude.

For $\delta = 4$, roughly $1\%$ of the total beam power in the atmosphere is deposited at densities higher than $10^{-11}\mathrm{g}/\mathrm{cm}^3$, in what might be considered chromospheric plasma. Adjusting the value of $\delta$ was found to roughly give a power-law variation in the percentage of non-thermal power deposited in the chromosphere, with significantly smaller percentages for higher values of $\delta$. However, the power-law exponent describing this relationship is highly dependent on the individual beam trajectory. For beam set 1, 2, and 3, the percentage of chromospheric power is respectively $10\%$, $1\%$, and $6\%$ for $\delta = 4$. The percentage is down-scaled by 1--3 orders of magnitude when going from $\delta = 3$ to $\delta = 6$.

\section{Discussion and conclusions}
\label{sec:discussion}
The key factor in determining the amount and energy of accelerated electrons in any part of the corona is the magnetic topology. Although a stronger magnetic field provides a larger source of energy, it is the magnetic topology that determines the potential for this energy to be released by reconnection. This is evident in the distribution of acceleration regions in our simulation. The upper corona, where the expanding magnetic bubbles collide with the weak overlying ambient field, contains acceleration regions that are equally energetic as acceleration regions in magnetically stronger, but topologically simpler parts of the lower corona. Consequently, the overall complexity of the magnetic field configuration is likely to be the main indicator of the significance of non-thermal energy transport.

In our simulation, the non-thermal power deposited at notable beam heating sites in the lower atmosphere range from $10^{18}$ to $10^{22}\;\mathrm{erg}/\mathrm{s}$, depending on the particular site and value of $\delta$. A typical small-scale beam heating event in the atmospheric conditions modelled here may then be estimated to release $10^{20}$--$10^{24}$ erg of non-thermal energy in the lower atmosphere, assuming the events to last $\sim 100$ s. Other heating mechanisms, including local Joule heating, thermal conduction, and magnetoacoustic shocks will make a significant additional contribution to the total energy release in some of these events \citep{Archontis2014, Hansteen2017, Hansteen2019}, one example being the heating near the strong central current sheet (beam set 3). Most of the beam heating events are nevertheless relatively weak, even for nanoflares. But they are highly abundant, and a $10 \times 10\;\mathrm{Mm}$ horizontal area of the chromosphere is likely to host a significant number of small beam heating events at any given time.

Even though the particle beams in this simulation are weak, their heating effect on the chromosphere is many orders of magnitude stronger than that of thermal conduction. This demonstrates that heating by energetic particles and thermal conduction in the lower atmosphere are qualitatively different, even under relatively quiet solar conditions. Because efficient thermal conduction requires a hot plasma, conductive transport always ceases at the bottom of the transition region. Incoming energetic particles, on the other hand, are not directly affected by the transition region temperature drop. The increase in mass density causes them to thermalise more quickly, but this occurs more gradually with depth than the abrupt shut-down of thermal conduction.

The inclusion of chromospheric heating by electron beams in atmospheric simulations such as the one used here could potentially account for discrepancies between synthetic diagnostics and observations. \citet{Testa2014} found that thermal conduction alone could not explain observed blueshifts of the SI IV spectral line in small-scale brightenings at coronal loop footpoints. Instead, their simulations, together with the extended analysis of \citet{Polito2018}, show that non-thermal electron beams can provide sufficient heating at the depths required to produce the upflows responsible for the blueshifts.

An advantage of considering the transport of accelerated particles in a 3D rather than a 1D atmospheric model is that it paints a realistic picture of how the available non-thermal energy is distributed in space. Although coherent large-scale flaring events may be reasonably approximated in a 1D coronal loop model with accelerated particles injected at the top, these types of idealised configurations are probably not representative of the situation in most active regions most of the time, and even less so outside of active regions. The quiet solar magnetic field tends to be tangled and inhomogeneous, which can lead to acceleration at any height in the corona and gives a complicated mapping from the acceleration regions to the locations where the non-thermal energy is deposited. Because acceleration takes place over extended regions in which the magnetic field changes topology, particles that are associated with the same reconnection event may end up on completely different trajectories through the atmosphere. Furthermore, the convergence of the magnetic field with depth can lead energetic particles that originate in separate acceleration regions to deposit their energy near the same location in the lower atmosphere (as exemplified by beam set 2 in Fig. \ref{fig:xz_power_change_selected_beams}).

The fact that beam heating sites can receive significant contributions of non-thermal electrons from several acceleration regions may have important observational consequences. The incoming electron beams could have been accelerated under different conditions, and thus do not necessarily have the same initial energy distributions. Moreover, beams coming from separate locations are influenced to varying degrees by collisions in the corona due to the different trajectories they take to the beam heating site. For instance, beams traversing a high column mass of coronal plasma lose a large share of their low-energy electrons, which gives them a hard energy distribution upon impact with the lower atmosphere. The total distribution of non-thermal electrons incident on the beam heating site could thus be a superposition of several distinct distributions. Consequently, the common assumption of a power-law distribution for the bremsstrahlung-emitting electrons in flares may not be applicable in all cases.

In future work, the response of the atmosphere to the electron beams will be investigated. It is also of interest to generate synthetic spectra from the beam heating sites. Furthermore, the development of the energetic particle model presented here is ongoing, and various improvements could be implemented.

Currently, our particle transport model does not include the effects of magnetic gradient forces. However, the strengthening of the magnetic field with depth in the chromosphere could significantly increase the pitch angle of the incoming electrons. If this effect is sufficiently strong, it will hamper the penetration of the most energetic electrons. Instead of thermalising below the photosphere, they might instead only reach the middle chromosphere, resulting in more beam heating at this depth. In general, a numerical treatment of the energy transport problem is required when considering magnetic gradient forces, although a simplified analytical approach has been suggested by \citet{Chandrashekar1986}.

The atmospheric simulation used for this paper assumes LTE and statistical equilibrium in its equation of state. The effects of non-equilibrium hydrogen ionisation are likely to alter the ionisation fraction and electron number densities in the chromosphere \citep{Leenaarts2007}, and could thus have a notable impact on the resulting distribution of beam heating. Moreover, because collisions with the non-thermal electrons can ionise neutral hydrogen atoms, the electron beams themselves contribute to an increase in the hydrogen ionisation rate \citep[see e.g.][]{Fang1993}. The resulting increase in the electron number density will, in turn, affect the chromospheric beam heating.

When enhanced collisional ionisation is taken into account, it may also be important to consider electrons accelerated below the transition region \citep[as done by][]{Fang2006}. Although the electrons are unable to transfer energy a significant distance away from the acceleration region due to the high plasma density, they still produce a local increase in the ionisation rate which would not be present if all the reconnection energy was converted directly into Joule heating.

\begin{acknowledgements}
      This research was supported by the Research Council of Norway, project number 250810, through its Centres of Excellence scheme, project number 262622, and through grants of computing time from the Programme for Supercomputing.
\end{acknowledgements}

\bibliographystyle{aa}
\bibliography{main.bib}

\end{document}